\documentclass{interact}

\usepackage{epstopdf}
\usepackage[caption=false]{subfig}

\usepackage[numbers,sort&compress]{natbib}
\bibpunct[, ]{[}{]}{,}{n}{,}{,}

\makeatletter
\def\NAT@def@citea{\def\@citea{\NAT@separator}}
\makeatother

\theoremstyle{plain}
\newtheorem{theorem}{Theorem}[section]
\newtheorem{lemma}[theorem]{Lemma}

\newtheorem{proposition}[theorem]{Proposition}

\theoremstyle{definition}

\theoremstyle{remark}
\newtheorem{remark}{Remark}

\usepackage[title]{appendix}
\usepackage{dcolumn}
\usepackage{amsmath,amsfonts,fancyhdr,hyperref,amsthm,color,graphicx, amssymb,
	stmaryrd,txfonts,siunitx,lscape,setspace, booktabs }
\usepackage{latexsym}
\usepackage{epsfig,overpic}\usepackage{amsmath}
\usepackage{amsfonts}
\usepackage{amssymb}

\newcommand{\xx}{\mathbf{x}}

\newcommand{\z}{\mathbf{z}}
\newcommand{\rr}{\mathbb{R}}

\newcommand{\m}{\mathbf{m}}
\newcommand{\p}{\mathbf{p}}

\renewcommand{\u}{\mathbf{u}}

\newcommand{\q}{\mathbf{q}}

\newcommand{\A}{\mathbf{A}}

\renewcommand{\P}{\mathbf{P}}
\newcommand{\be}{\begin{equation}}
\newcommand{\ee}{\end{equation}}

\newcommand{\ve}{\varepsilon}

\newcommand{\n}{\mathbf{n}}

\renewcommand{\v}{\mathbf{v}}

\newcommand{\E}{\mathbf{E}}

\newcommand{\tr}{\text{tr }}
\newcommand{\partderiv}[2]{\ensuremath{\frac{\partial #1}{\partial #2}}}

\newcommand{\Tau}{T}

\title{Three-dimensional soliton-like distortions in flexoelectric nematic liquid crystals:
modeling and linear analysis}
\author{Ashley Earls\textsuperscript{a} and M.~Carme Calderer\textsuperscript{b}\thanks{CONTACT M.C.Calderer. Email: calde014@umn.edu} \affil{\textsuperscript{a}Basc Center for Applied Mathematics, Alameda de Mazarredo 14,	48009 Bilbao, Bizkaia,	Basque-Country, Spain \textsuperscript{b}School of Mathematics, 
		University of Minnesota,
		Minneapolis, MN 55455, USA }
}

\begin{document}
\maketitle
\begin{abstract}{
This article models  experimentally observed three dimensional particle-like waves that develop in nematic liquid crystals, with negative dielectric and conductive anisotropy, when 
subject to an applied alternating electric field. The liquid crystal is  confined in a thin region between two plates, perpendicular to the applied field. The horizontal, uniformly aligned 
director field is at equilibrium due to the negative anisotropy of the media. However, such a state is unstable to perturbations that manifest themselves as confined, bullet-like, director
distortions traveling up and down the sample at a speed of several hundred microns per second.  It is experimentally predicted that flexoelectricity plays a key role in generating the soliton-like behavior. We develop a variational  model that accounts for ansiostropic dielectric, conductive, flexolectric, elastic and  viscous forces. We perform a stability analysis of the uniformly aligned equilibrium state to determine the threshold wave numbers, size, phase-shift and  speed of  the soliton-like disturbance. We show that the model predictions  are in very good agreement with the experimentally measured values.  
The work models and analyzes a  three-dimensional soliton-like  instability reported, for the first time  in flexoelectric liquid crystals, pointing towards a potential application as a new type of nanotransport device.
}\end{abstract}

\begin{keywords} {negative dielectric and conducting anisotropy; soliton-like wave; flexoelectricity; unstability}
	\end{keywords}
\begin{amscode}  37N10, 34C11, 34C29, 34D20 \end{amscode}
\begin{pacscode}  02.30.Hg, 02.30.Jr, 02.30.Nw, 05.45.Yv\end{pacscode}






\maketitle
\section{Introduction}

In this paper, we study a new type of three-dimensional soliton-like director field  distortion observed  in a flexoelectric nematic liquid crystals with negative dielectric anisotropy, subject to an (AC) alternating electric field \cite{Tuxedos}. The traveling disturbances, referred to as {\it director bullets},  {\it tuxedos} and {\it soliton-like} structures,   consist of  concentrated three-dimensional pattern of chevron-like distortion of molecular alignment of the liquid crystal  propagating through a uniformly oriented sample. They arise as  perturbations of the director field from the uniform state.
Within the bullet  region,  the bow-like director perturbation oscillates with the frequency of the applied AC electric field and breaks the fore-aft symmetry,  resulting in rapid propagation perpendicularly to the initial direction of alignment. They do not spread while moving over  macroscopic distances a  thousand times longer than their size.   In this article, we build a mathematical model of the phenomenon that accounts for the intertwined effects of  negative dielectric and electric charge conducting anisotropy  together with the flexoelectric, bending and viscous, contributions, all finely tuned, as collectively respond to the applied AC-field.  We analyze the instability threshold of the undisturbed, uniformly alignd base solution and show that it   characterizes the size and speed of the emerging soliton-like disturbance.  
The governing system couples the Ericksen-Leslie equations of the dynamics of the director field $\n$, including dielectric and flexoeletric effects,  but neglecting defects and  flow, coupled with the Poisson-Nearst-Planck system of electric charge motion.

\begin{figure} \label{geometry}
\centering
\includegraphics[height=0.9in]{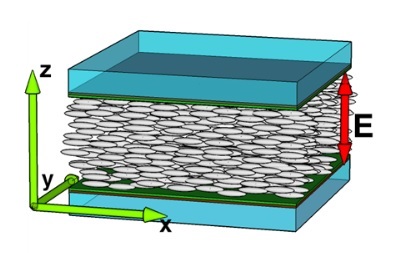}\hspace{.1in}
\includegraphics[height=0.9in]{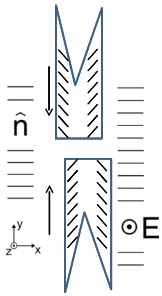}\hspace{.1in}
\includegraphics[scale=.5]{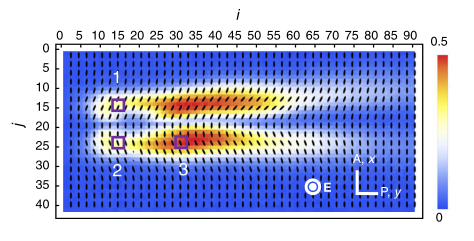}
\\ \vspace{.1 in}
\includegraphics[height=1.0in]{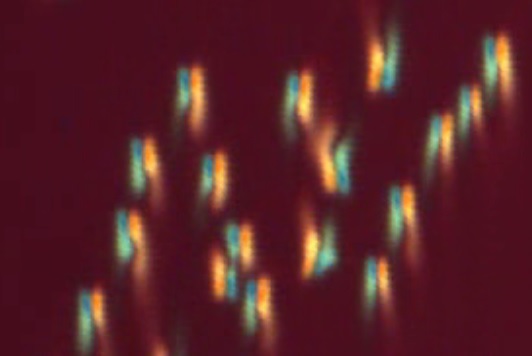}
\includegraphics[scale=.4]{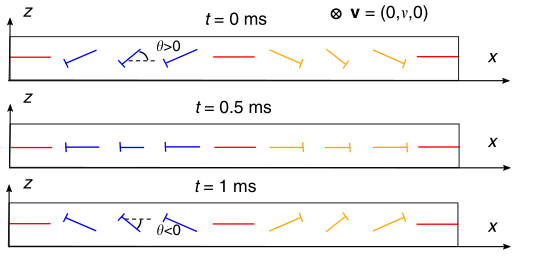}
\caption{Top left: Geometry and initial alignment of the director field \cite{Oleg}. Middle sketch: schematic of the tuxedo structure and  snapshot of  tuxedo moving left in relation to the page  \cite{Oleg} (top right). The direction field shows the angle $\phi$ and the color is the measured light intensity. The units are $\SI{1}{\micro\meter}$ along each axis. The planar angle $\phi$ is largest at location $3$, and $\theta$ is zero in the chevron except at locations $1$ and $2$ \cite{Tuxedos}. Bottom left: tuxedos traveling up and down in relation to the page \cite{Oleg}. This is a top-down view of the middle plane between the two plates, at $z=0$.  
Bottom right: A schematic illustrating the angle $\theta$ when the period of the external electric field is $\SI{2}{\milli\second}$. The nails indicate the director $\n$, with the heads closer to the reader than the ends. The vector $\v$ indicates the velocity of the chevron \cite{Tuxedos}. }
\label{fig1}
\end{figure}

This paper models the experiments reported in \cite{Tuxedos}, where a sample of 4prime-butyl-4-heptyl-bicyclohexyl-4-carbonitrile (CCN-47) with impurities is confined  between two parallel plates, located at $z=0$ and $z=d$, respectively, of a Cartesian coordinate system $(x, y, z)$ of the Eucledian space. The plates form a cell of thickness $d=3-\SI{30}{\micro\meter}$ (figure \ref{fig1}, left picture).  Initially, the director is uniformly aligned parallel to the plates, $\n_0=\hat{\xx}$, and an alternating  electric field is then applied across the cell in the perpendicular direction to the plates, $\E_0=E_0\cos\omega t\;\hat{\z}$. Note that  $\n_0$  is an equilibrium state  of the system due to the negative dielectric anisotropy of the media, with the property that molecules tend to align on a plane perpendicular to the applied electric field. The amplitude of the applied voltage is within the range $U=10-\SI{90}{\volt}$. Although  a wide  range of frequencies is being used in the experiment, $\omega = 20-\SI{5000}{Hz}$, only the smaller interval between $400$ and $\SI{460}{Hz}$ is reported to generate soliton-like disturbance. Once generated, they travel up and down the region on the direction, taken to be the $y$-coordinate axis, perpendicular to both   $\n_0$ and $\E_0$. 
The bullets exhibit all the hallmarks of soliton behavior: (1) they move with constant speed, (2) do not decay or disperse, and (3) retain their shape after pairwise collisions.  Moreover, the disturbances form only in the middle plane of the cell, $z=d/2$, away from the bounding plates,  and therefore are truly three-dimensional. Neither defects nor net flow are observed in the sample. To describe their shape, let us consider a single chevron and take the $x$-axis to be along its center.
The orthonormal set of unit vectors $(\n, \p, \m)$ play a main role in setting the geometry of the soliton-like package:
\be
\n = \left(\cos\theta\cos\phi,\cos\theta\sin\phi,\sin\theta\right), \quad 
\m= \left(\sin\phi,-\cos\phi\right), \quad \p=\n\times\m \label{nParam} \ee
with $\n$ representing the parametrization of the director field in terms of the angles $\theta$, out-of-plane,  and $\phi$, the in-(x,y) plane angle giving the disturbance a chevron form. 

We develop a variational  model that accounts for dielectric, conductive, flexolectric, elastic and anisotropic viscous forces.
It consists of a  nonlinear parabolic system of partial differential equations for the  dynamics of the director field $\n$ and  the Poisson-Nernst-Planck system for the diffusion and transport of electric charge, with $c^+$ and $c^-$ representing the concentrations of positive and negative ions, respectively. These equations are coupled with Poisson's equation for 
the electrostatic potential $\Phi$. The equations of the director dynamics follow from the Ericksen-Leslie equations of liquid crystal flow setting the velocity field equal to zero, according to the experimental observation of no net flow taking place. 
The  dimensional analysis of the governing equations reveals four time scales, that listed in increasing order include: the time scale of the dielectric effects $T_d$, the flexoelectric time scale $T_f$ that is comparable in magnitude to that of the applied AC field $T_\omega$ and finally that of the elastic effects, $T_e$.  
Furthermore, the time scale $T_d$ is  associated with the initial layer behavior of the dielectric terms, acting at the beginning of every AC-cycle, approximately during the first 0.01 dimensionless time units. This allows us to obtain an approximate governing system,  past the initial time, that excludes the appropriate dielectric terms. 
We also show that solutions of the governing system have the experimentally observed  symmetry properties, and with every bullet traveling down along the y-axis, there is a symmetric one moving on the opposite direction. The predictions of the $\frac{\pi}{2}$ phase-shift of the out-of-plane angel $\theta$ is  accurate within $10^{-2}$ error.

Our work focuses on characterizing the instability threshold of the uniform solution, $\n_0$, and explore the corresponding conditions to determine the size range, phase and speed of the subsequent disturbance.   For this, we linearize the governing system about the basic state $\n_0$ and take the double space Fourier transform  of the resulting equations. 
The latter consists of a coupled system of three ordinary differential equations, with respect to the time variable $t$, for the angles $\phi$ and $\theta$ and the net charge $q$ (the equation for the total charge, $Q$, decouples from the rest).  The equations contain the parameters of the problem together with the (justified as purely imaginary) wave numbers, $\rho_x$ and $\rho_\xi$, of the perturbation, horizontal and vertical, respectively. (The vertical wave number refers to the traveling wave variable, $\xi:=y-Rt$, $R$ being the dimensionless speed of the perturbation). We perform two main types of simplification on the system, first, we average it with respect to the cross-sectional variable $z$, with an ansatz motivated by a polynomial expansion of the unknown fields with respect to $z$, that captures the boundary conditions on top and bottom plates. The second simplifying assumption consits on a time averaging of the system with respect to the small parameter $L_i^{2}$, where $L_i$ is a dimensionless constant measuring a relative strength of the flexoelectric effects (of the order $O(10^{-3})$).    
Finally, the application of the variation of constants formula allows us to obtain the solution of the (approximated) governing system. We show that the uniform solution is an unstable node of the system, which together with the expressions of the general solutions allow us to formulate the conditions on the eigenvalues of the system that lead to (the approximate) neutral stability of the equilibrium state.  These turn out to be, in part,  conditions on the trace of  two matrices, very much along the lines of the Floquet theory, leading to satisfactory bounds on the size of the disturbance. Specifically, the bound on the horizontal size, that of twice the distance between the two plates, is accurate, the vertical length of the disturbance as predicted by the model falls on the lower range of the experimentally measured soliton-like lengths, that is, between $2.5\times d$ and $6.25\times d$ $\mu$m, where $d=8\mu$. The model  underestimates the speed of the observed disturbance, by about a factor of $\frac{1}{2}$, also with respect to the lower bound values of experimental measurements. The linearity of the model may be responsible for the underpredictions. 
Our analysis also shows that,  instabilities do not occur without flexoelectric polarization in the model, and neither occur in the absence of ions \cite{buka2013flexoelectricity}.
Moreover, we find  that the presence of ionic impurities as well as the increase of the absolute value of the anisotropic conductivity, each  contributes to increasing  the speed of the bullets.  The shape of the chevron bullet is described in figure \ref{fig1}.
The  pattern formation by application of an electric field to a liquid crystal  has a long experimental and modeling history
\cite{buka2006convective}, \cite{dGProst}. These patterns can be widely classified into two types, director distortions 
associated with the Freedericks transition \cite{TF9332900919}, \cite{zocher1933effect} and electroconvective, along the vein of well-known phenomena such as the Taylor-B{\'e}nard convection \cite{benard1900tourbillons}, \cite{rayleigh1916lix}, \cite{koschmieder1993benard}.
They  are  rooted in two key 
properties, the  dielectric and conductive anisotropies, $\ve_a$ and $\sigma_a$,  respectively, in addition to the geometry, either planar or homeotropic,  of the undistorted director alignment $\boldsymbol n_0$. 
We recall that in materials with $\ve_a>0$, the liquid crystal tends
to align with the electric field, whereas the alignment is transverse in the negative case. 
Likewise, ionic impurities will move along the direction of the field for liquid crystals with $\sigma_a>0$,
and transversely otherwise.

Since the pioneering  works by Williams \cite{williams1963domains}  and Kaspustin and Larinova \cite{kapustin1964behavior}, there have been extensive experimental and theoretical investigations of the electro-hydrodynamic convection in nematic liquid crystals. In William's experiment, the nematic, characterized by $\epsilon_a<0$ and $\sigma_a>0$,  is enclosed between two parallel plates, with separation $d$ between 10-100$\mu$, in a geometry identical to the one in this work, with director alignment $\n_0= \hat{\xx}$ (Fig. 1, top left). 
Applying a low frequency AC field, with a threshold maximum voltage of 10 V, convection rolls  appear with associated director distortions, easily detected optically. 
These are rolls along the $y$-direction parallelly stacked next to each other, at a distance $d$, along the $x$-axis.  Carr \cite{carr1969influence},
		provided the theoretical bases of the Williams domains, starting with the observation that the motion of   ions  along the electric field direction  causes  $\n$ to tilt towards  the $z$ axis. Elastic forces resist the tilting, leading to a periodic equilibrium configuration with spatial concentrations of charge, especially in regions where the director gradient is large. Finally, the localized charge coupled with the electric field induces a circular flow  reminiscent of Rayleigh-B\'{e}rnard convection, with the same periodicity as the director field. This pattern persists even when the electric field changes sign, and as shown in \cite{DuboisViolette}, the critical voltage at which the Williams domains occur is independent of the sample thickness $d$. 
		 The first theoretical study of the Carr effect, in one-dimensional geometry,  was carried out by Helfrich \cite{helfrich1969conduction} followed by two-dimensional analyses   of the phenomenon\cite{dubois1971hydrodynamic}. The full three-dimensional analysis as well as the pattern formation for large frequencies (dielectric regime) and more general geometries was carried out by Kramer and Pesch \cite{kramer1995convection}. Experiments and analysis of several liquid crystals revealed the three main features that characterize electroconvection: geometry (either planar or homeotropic alignment) and the signs of $\epsilon_a$ and $\sigma_a$. Liquid crystals such that $\epsilon_a\sigma_a<0$ and, for both initial director field orientations, present primary convective pattern. However those such that $\epsilon_a<0$ and $\sigma_a<0$, with planar initial director orientation, fall into the class known as  presenting non-standard Carr-Helfrich mechanism, with rarely observed convection phenomenon; they are still the subject of active investigation \cite{wiant2005nonstandard,  shiomi2020prewavy, huh2021reentrant}. This is the case of the material in our study, and for which electroconvective phenomena has not been observed.

A related effect that does not involve ionic impurities comes from the often neglected flexoelectric polarization, that is, the development of electric dipoles associated with director field gradients \cite{meyer1969piezoelectric}.  Each rod-shaped molecule of the nematic carries a small dipole moment, and in a nearly uniformly-aligned sample, these dipoles nearly cancel and therefore have little effect. However, when the sample is subject to splay or bend type distortions, the small charges present on each molecule accumulate, leading to a spatial separation of charge. Flexolectricity is also known to generate rich equilibrium spatial pattern \cite{krekhov2011flexoelectricity}. 
 Whereas Williams domains are entirely explained by ionic charge,  soliton-packages cannot occur without flexoelectric polarization, as observed in \cite{Tuxedos}.

Soliton-like waves are  ubiquitous to many physical systems and have been extensively studied \cite{zabusky1965interaction}. Mathematically, solitons refer to solutions of distinctive  partial differential equations associated with a Hamiltonian system, that in spite of being of dispersive type,  present the  three wave-particle features previously described. Although solitons are regarded as stable solutions of the PDE due to the long time persistence of the shape, the question of their stability, specially for PDEs with 
critical nonlinearity, remains one of  the most challenging open problems  of mathematical physics \cite{tao2009solitons}.
Equations of this class include the KdV equation, the wave equation, and the linear and nonlinear Schr\"{o}dinger equations.
Optical solitons in liquid crystals, {\it nematicons},  forming in nonliner optic regimes prevalent, for instance,  in applications to fiber optics,
are associated with solutions of the latter \cite{Minzonie,Borgna}. The type of phenomena studied in this article falls into the class of particle-like waves known as {\it physical solitons}.
The goal of the experimental work is to design new mechanisms of particle nanotrasport. 
The article is organized as follows. In section 2, we derive the governing equations, obtain the relevant nondimensional parameter groups and time-scales, in particular, leading to the identification of the initial layer term.  In section 3, we linearize the PDE system about the initial uniform state and apply Fourier analysis and asymptotic analysis to obtain a sufficient condition for instability.  Concluding remarks are given in section 5.
This work is part of the Ph.D. thesis dissertation by Ashley Earls \cite{Ashely}.

\section{A model}  The goal of this section is to derive the set of governing equations appropriate to describe the soliton-like phenomenon that we aim at investigating. We assume that the liquid crystal occupies a domain $\Omega\subset {\mathbb R}^3$ that will later take to be the paralepiped region confined between the two electrodes. The proposed model  involves the main physical ingredients of the phenomenon: dielectric and flexoelectric polarization, elastic,  viscous and ionic effects, and the role of the alternating electric field.  It also takes into account two main simplifying assumptions, one of them being the absence of an explicit role of defects in the evolution of the chevron structures, which are interpreted as traveling distortions of the molecular alignment that do not generate net flow. This allows us to describe the state of alignment of the liquid crystal by the unit director field $\n$, and 
follow a modeling approach based on the Ericksen-Leslie equations, suppressing flow but including electrodynamic effects and flexoelectric polarization. 
 Furthermore, rather than stating the laws of balance of linear and angular momentum of the system, we follow the variational approach associated with the principle of minimum dissipation.  For this, we proceed to writing the total energy and the rate of dissipation function of the system, for the set of generalized variables $\mathbf q$ and generalized velocities $\dot \q$.  In addition to the director $\n$, $\mathbf q$ includes the family of charge concentration variables $c^k$, $1\leq k\leq N$, corresponding to $N$  ionic  species with valence $z^k$, and the electrostatic potential $\Phi$. The generalized velocities consist of $\dot \n$, $\dot\Phi$ and $\u^k$, $1\leq k\leq N$, the latter denoting the ionic velocity of the $k$th species.  Let $c_0>0$ denote a typical value of the concentration of one of the ionic families in the problem. The value of $c_0$ specified in the table is that given in \cite{Tuxedos}. Approaches to precise measurements of such a quantity in liquid crystal cells have been recently reported in the literature \cite{garbovskiy2021evaluating}. The ion concentrations satisfy the continuity equations 
 \be
 \frac{\partial c^k}{\partial t}+\nabla\cdot(c^k\u^k)=0, \quad 1\leq k\leq N.
 \label{eqn:continuity0}
 \ee 
The anisotropy of the liquid crystal is encoded in its dielectric tensor 
 \begin{equation}
 \boldsymbol{\ve}(\n)= \ve_0(\ve_\perp I + \ve_a\n\otimes \n), 
 \quad \ve_a:=\ve_\|-
 \ve_\perp,
 \end{equation}
 where $\ve_0$ is the vacuum dielectric constant and $\ve_a$ denotes the dielectric anisotropy, the difference between the material dependent parallel and perpendicular anisotropy coefficients. 
 Flexoelectric materials are characterized by  the electric displacement vector field
 \begin{equation}
 \mathbf D_\text{elec}= \ve(\n)\mathbf E + \mathbf P_\text{flex},
 \end{equation}
 where $\mathbf P_\text{flex}$ denotes the polarization field of the material, that is, the dipolar density due to splay and bend distortions of the director field \cite{dGProst},
 \begin{equation}
 \P_{\text{flex}} = e_1(\nabla\cdot\n)\n+e_3(\nabla\times\n)\times\n = e_1(\nabla\cdot\n)\n+e_3(\nabla\n)\n.
 \label{PflexDef}
 \end{equation}
The coefficients $e_1$ and $e_3$ in \eqref{PflexDef} can be positive or negative, and the second equality in \eqref{PflexDef} follows from the identity $\v\times(\nabla\times\v)=-(\nabla\v)\v$ for $|\v|=1$.
\noindent The total energy density of the system consists of the contributions
\begin{equation}
\mathcal{E}=\mathcal{E}_\text{OF}+\mathcal{E}_\text{dielc}+\mathcal{E}_\text{ion}+\mathcal{E}_\text{flex} =:\tilde{\mathcal{E}}+\mathcal{E}_\text{flex},
\end{equation}
$\mathcal{E}_\text{OF}$ denoting the Oseen-Frank energy of director distortion, the dielectric electrostatic  energy  $\mathcal{E}_\text{dielc}$, the ionic contribution  $\mathcal{E}_\text{ion}$ which involves the entropy
as well as the electrostatic energy of free charged particles, and the electrostatic contribution of the flexoelectric polarization $\mathcal{E}_\text{flex}$. These are of the form
\be\begin{aligned}
& \mathcal{E}_\text{OF} = \tfrac{1}{2}K_1(\nabla\cdot\n)^2+\tfrac{1}{2}K_2[\n\cdot(\nabla\times\n)]^2+\tfrac{1}{2}K_3|\n\times(\nabla\times\n)|^2+ \tfrac{1}{2}(K_2-K_4)\left[\tr^2(\nabla\n)-\tr({\nabla 
\n}^2)\right],\\
& \mathcal{E}_\text{dielc}=-\tfrac{1}{2}\ve_0\left[|\E|^2+\ve_a(\n\cdot\E)^2\right],\quad 
\mathcal{E}_\text{ion} = e\Phi\sum_{k=1}^Nz^kc^k+K_B\Tau\sum_{k=1}^Nc^k\ln\left({\frac{c^k}{c_0}}\right),\quad 
\mathcal{E}_\text{flex} = -\mathbf{P}_\text{flex}\cdot\E,
\end{aligned}\ee
where the positive constants  $\Tau$ and $e$ denote the absolute temperature and  the elementary charge, respectively, and $K_B$ the Boltzman constant.  The Frank elasticity constants $K_i$ satisfy,
$K_1>0, \, K_2>0,\, K_3>0,$  $ \, K_2\geq |K_4|, \, 2K_1\geq K_2+K_4.$ 

We briefly recall that the dynamics of a non-dissipative system, in a domain $\Omega\subset {\mathbb R}^3$,  described by the generalized coordinates $q_i$ and the generalized velocities $\dot q_i$,  is given by the equations
\be
X^e_i:=\frac{d}{dt}\frac{\partial\mathcal{L}}{\partial\dot{q_i}}-\frac{\partial\mathcal{L}}{\partial q_i}=0,\qquad \mathcal{L}=\mathcal{T}(\dot{q}_i)-\mathcal{U}(q).
\ee
Here $\mathcal{L}$ denotes the Lagrangian of the system, that is, the difference of the density of kinetic energy $\mathcal{T}(\dot{q}_i$), the density of potential energy $\mathcal{U}(q_i)$,  and $X_i^e$ represents the elastic force \cite{Sonnet2001}. The variational statement of this equation is
$
\frac{\delta}{\delta \dot{q}_i}\int_{\mathcal P} \dot{\mathcal{E}}\, d\xx=0,
$
where $\mathcal{E}=\mathcal{L}+2\mathcal{U}$ is the total energy of the system and $\mathcal P\subseteq \Omega$.  That is, the system behaves in such a way that the rate of work is minimized with respect to the generalized velocities. 
Letting $\mathcal{R}=\mathcal R(q_i, \dot q_i)$ represent the Rayleigh dissipation function, the dissipative forces are  given by $X_i^d=-\partial\mathcal R/\partial\dot q_i$.  The dynamics of a dissipative system is then formulated as the balance of the conservative forces by the dissipative ones, that is, the statement 
\be
\frac{d}{dt}\frac{\partial\mathcal{L}}{\partial\dot{q_i}}-\frac{\partial\mathcal{L}}{\partial q_i}+\frac{\partial\mathcal{R}}{\partial\dot{q}_i}=0.
\ee
 As in the conservative case, the above equations also have a variational interpretation in that they are critical points
of the Rayleigian functional with respect to $\dot\q$:
\be
\frac{\delta}{\delta\dot{q}_i}\int_{\mathcal P} (\dot{\mathcal{E}}+\mathcal{R}-\chi \n\cdot\dot \n)\, d\xx=0,\,\label{PMD1}
\ee
with the variations performed while holding $\q$ and the elastic forces $X_i^e$ constant. The last term in the previous equation corresponds to imposing the unit director constraint $|\n|=1$, with $\chi $ denoting the Lagrange multiplier. 
The two sources of energy dissipation of the system are the rotational viscosity of the director field
and the diffusion of the ionic particles,
\begin{equation}
\mathcal R= 
\frac{1}{2}\gamma_1|\dot\n|^2 +  \frac{K_B\Tau}{2}\sum_{k=1}^Nc^k\u^k\cdot\mathcal{D}^{-1}\u^k,
\end{equation} 
where $\u^k$ denotes the velocity of the $k$th ionic species (with respect to the nematic fluid, at zero velocity) and $\mathcal D$ the diffusion matrix, taken to be the same for all ion types. The principle of minimum dissipation (\ref{PMD1}) yields the relations
\begin{align}
0 &= \frac{\delta}{\delta \dot\n}\int_\Omega\left(\dot{\tilde{\mathcal{E}}}+\mathcal{R}-\chi\n\cdot\dot\n\right)\nonumber\\
&= \int_\Omega\left[\frac{\partial}{\partial\n}(\mathcal{E}_\text{OF}+ \mathcal{E}_\text{flex}) -\nabla\cdot\left(\frac{\partial}{\partial\nabla\n} (\mathcal{E}_\text{OF}+\mathcal{E}_\text{flex})\right)-\chi\n+\gamma_1\dot{\n}-\ve_0\ve_a(\n\cdot\E)\E\right],\label{paperResults1}\\
0 &= \frac{\delta}{\delta \dot\Phi}\int_\Omega\left(\dot{\tilde{\mathcal{E}}}+\mathcal{R}-\chi\n\cdot\dot\n\right)=\int_\Omega \left[-\nabla\cdot\left[\ve_0\left(\ve_\perp\mathbf{I}+\ve_a\n\otimes\n\right)\E\right]+e\sum_{k=1}^cN^kz^k-\nabla\cdot\P_\text{flex}\right],
\label{paperResults4}\\
0 &= \frac{\delta}{\delta \u^k}\int_\Omega\left(\dot{\tilde{\mathcal{E}}}+\mathcal{R}-\chi\n\cdot\dot\n\right)=\int_\Omega c^k\left[\nabla\mu^k+K_B\Tau\mathcal{D}^{-1}\u^k\right].\label{paperResults3}
\end{align}

These equations are complemented by the natural boundary conditions. 
 The term $\mu^k$ in \eqref{paperResults3} denotes  the chemical potential associated with ion species $k$, given by
\begin{equation}\mu^k=\partderiv{\mathcal E_{\text{ion}}}{c^k}=e\Phi z^j +K_B\Tau\left[\ln \left(\frac{c^k}{c_0}\right)+1\right], \quad 1\leq k\leq N. \label{eqn:chem-potential} \end{equation}
Therefore \eqref{paperResults3} yields
$
\u^k = -\frac{1}{K_B\Tau}\mathcal D\nabla\mu^k = -\mathcal{D}\left(\frac{\nabla c^k}{c^k}-\frac{ez^k}{K_B\Tau}\E\right). $
Substituting the latter into (\ref{eqn:continuity0}) 
 and assuming that (\ref{paperResults1}) and (\ref{paperResults4}) hold for every subpart $\mathcal P\subseteq\Omega$, we arrive at the governing system of partial differential equations and the constraint relations:
\be \left\{\begin{aligned}
& \frac{\partial\mathcal{E}_\text{OF}}{\partial\n}-\text{div}\left(\frac{\partial\mathcal{E}_\text{OF}}{\partial\nabla\n}\right)-\chi\n
+\gamma_1\dot{\n}-\ve_0\ve_a(\n\cdot\E)\E\\
&\hspace{1in}
+(e_3-e_1)[(\nabla\cdot\n)\E-\nabla\n^\intercal\E]+(e_1+e_3)(\nabla\E)\n=\mathbf{0},\\
& \nabla\cdot\left[\ve_0\ve_\perp\E+\ve_0\ve_a(\n\cdot\E)\n
+e_1(\nabla\cdot\n)\n+e_3(\nabla\n)\n\right]=e\sum_{k=1}^Nz^kc^k,\\
& \frac{\partial c^k}{\partial t}+\nabla\cdot\left[\mathcal{D}\left(\frac{ez^kc^k}{K_B\Tau}\E-\nabla c^k\right)\right]=0,\\
& \n\cdot\n = 1,\quad 
 \E=-\nabla\Phi, \quad \Phi \text{denotes the electrostatic potential}. 
\end{aligned}\right.\label{MCCeqns}\ee

\subsection{The governing equations of the chevron system.}
\noindent The experimental domain is the liquid crystal region $\Omega$ enclosed between two parallel plates 
\begin{equation} \Omega=\left\{(x, y, z)\in {\mathbb R}^3: -\frac{L}{2}<x<\frac{L}{2}, \, 0<y<L, \, 0<x_3<d\right\}:=\Omega_\perp\times (0, d). \label{domain}\end{equation}
\noindent To describe the chevrons reported in \cite{Tuxedos}, we make the following assumptions:
\begin{enumerate}
\item The simplifying, one-constant approximation of  the Oseen-Frank energy, with $K_1=K_2=K_3=K$ and $K_4=0$. 
\item There are two ion species present in the sample ($N=2$), with concentrations $c^\pm$ and valences $z^\pm=\pm 1$.
\item The dielectric anisotropy is negative: $\ve_a<0$.
\item The diffusion matrix for the ions is given by
\be
\mathcal{D} = \bar{D}\left[\mathbf{I}+(\lambda_\sigma-1)\n\otimes\n\right],\qquad \bar{D}=\frac{K_B\Tau}{e^2}\frac{\sigma_\perp}{c_0}. \label{Dbar}
\ee

\item The conductive anisotropies satisfy $\sigma_\parallel<\sigma_\perp$, so $\lambda_\sigma:=\sigma_\parallel/\sigma_\perp<1$.

\end{enumerate}
 Using the Euler representations \eqref{nParam}  of the unit vectors $\{\n,\m,\p\}$, the governing equations \eqref{MCCeqns} become
\be\left\{\begin{aligned}
& \gamma_1\phi_t = K\left(\Delta\phi-2\tan\theta\;\nabla\phi\cdot\nabla\theta\right)
+\ve_0|\ve_a|(\n\cdot\E)(\m\cdot\E)\sec\theta+(e_1+e_3)(\nabla\E)\n\cdot\m\\
&
\hspace{1in}+(e_1-e_3)\left[(\nabla\theta\cdot\p)(\E\cdot\m)-(\nabla\theta\cdot\m)(\E\cdot\p)\right]\sec\theta, \\
&\gamma_1\theta_t = K\left(\Delta\theta+\tfrac{1}{2}\sin2\theta\;|\nabla\phi|^2\right)
+\ve_0|\ve_a|(\n\cdot\E)(\p\cdot\E)+(e_1+e_3)(\nabla\E)\n\cdot\p\\
& \hspace{1in}+(e_1-e_3)\cos\theta\left[(\nabla\phi\cdot\m)(\E\cdot\p)-(\nabla\phi\cdot\p)(\E\cdot\m)\right],\\
&\nabla\cdot\left[\ve_0\ve_\perp\E-\ve_0|\ve_a|(\n\cdot\E)\n+e_1(\nabla\cdot\n)\n+e_3(\nabla\n)\n\right]=e(c^+-c^-),\quad \E=-\nabla\Phi\\
&c^\pm_t = \bar{D}\nabla\cdot\left[(\mathbf{I}+(\lambda_\sigma-1)\n\otimes\n)\left(\nabla c^\pm\mp \frac{c^\pm e}{K_B\Tau}\E\right)\right]. 
\end{aligned}\right.\label{224}\ee
The unknown fields of the system are $\phi$, $\theta$, $\Phi$, $c^+$, and $c^-$, with domain in $\Omega$ and with 
$t>0$.
These  equations form a coupled nonlinear parabolic-elliptic system with an electric field source term. 
\subsection{Scaling analysis and nondimensionalization}
We define the  dimensionless variables
\begin{equation}
\bar x=\frac{x}{L}, \quad \bar y=\frac{y}{L}, \quad \bar z=\frac{z}{d}, \quad \eta=\frac{d}{L}, \quad \bar t=\frac{t}{T}, \quad  
\bar{\Phi}=\frac{\Phi}{E_0d},\quad
\bar c^\pm=\frac{c^\pm}{c_0},
\end{equation}
where $T>0$ denotes a time scale to be chosen later and $\eta>0$ is the aspect ratio of the domain. Upon scaling, the domain  $\Omega$ and its cross-section $\Omega_\perp$ in (\ref{domain}) become $\bar \Omega$
 and $\bar\Omega_\perp$, respectively.  The scalar $E_0>0$ denotes the maximum intensity of the applied AC field, with the product $E_0 d$ representing the value of the electric potential applied to the system. Likewise, $c_0>0$ denotes the typical concentration of a representative ionic species in the cell.  
\begin{small}\begin{center}{ 
 \begin{table}
\label{param_table}
\def\arraystretch{1.5}
\begin{tabular}{|l|l|l|}
\hline
Parameter & Label\hfill & Value (SI) \\ \hline
Plate size & $L$ & $\SI{5}{\milli\meter} $  \\
Plate separation & $d$ & $3-\SI{19.5}{\micro\meter}$  \\
Frank elastic constant & $K$ & $\SI{e-11}{\newton}$  \\
Rotational viscosity & $\gamma_1$ &  $\SI{6e-2}{\pascal\second}$  \\
Temperature & $\Tau$ & $\SI{313}{\kelvin}$   \\
Electric field intensity & $E_0$ & $\SI {8.2e6}{\volt\per\meter}$ \\
Electric field frequency & $\omega$ & $400-\SI{450}{\hertz}$   \\
Dielectric permittivity vacuum & $\ve_0$ & $\SI{8.85e-12}{\farad\per\meter}$  \\
Dielectric anisotropy & $\ve_a$ & $-4.2$ \\
Dieletric permittivity $\perp$ & $\ve_\perp$ & $8.8$ \\
Anisotropic conductivity $\parallel$ & $\sigma_\parallel$ & $\SI{4.9e-9}{\per\ohm\per\meter}$  \\
Anisotropic conductivity $\perp$ & $\sigma_\perp$ & $\SI{6.1e-9}{\per\ohm\per\meter}$ \\
Flexoelectric coefficients & $e_1$, $e_3$ & on the order of $\SI{e-11}{\coulomb\per\meter}$ \\
Typical charge concentration & $c_0$ & $\SI{2e20}{\per\meter\cubed}$\\
\hline
\end{tabular}
\label{table-parameter}
\end{table}}\end{center}\end{small} 
 The dimensionless version of the  system \eqref{224} is
\begin{align}
& B\phi_t
= 
C\left(\bar\Delta\phi-2\tan\theta\;\bar\nabla\phi\cdot
\bar\nabla\theta\right)+(\n\cdot{\bar\E})(\m\cdot{\bar\E})\sec\theta
+ L_2\sec\theta\left[\m\cdot(\bar\nabla{\bar\E})\n\right] \nonumber\\
&\qquad\qquad
+ L_1\left[(\bar\nabla\theta\cdot\p)(\bar\E\cdot\m)-(\bar\nabla\theta\cdot\m)(\bar\E\cdot\p)\right]\sec\theta, \label{eqn:dimensionless0-1}\\
& B\theta_t
= C\left(\bar\Delta\theta+\tfrac{1}{2}\sin2\theta\;|\bar\nabla\phi|^2\right)
+(\n\cdot\bar\E)(\bar\E\cdot\p)+ L_2\left[\p\cdot(\bar \nabla\bar \E)\n\right] \nonumber\\
&\qquad\qquad
+ L_1\cos\theta\left[(\bar\nabla\phi\cdot\m)(\bar\E\cdot\p)-(\bar\nabla\phi\cdot\p)(\bar\E\cdot\m)\right],\label{eqn:dimensionless0-2}\\
& \bar\nabla\cdot\left[J\bar\E-(\n\cdot\bar\E)\n+( L_1+ L_2)(\bar\nabla\cdot\n)\n+( L_1- L_2)(\bar\nabla\n)\n\right]=M(c^+-c^-), \quad \bar{\E}=-\bar{\nabla}\Phi \label{eqn:dimensionless0-3} \\
& Fc^\pm_t= \bar\nabla\cdot\left[(\mathbf{I}+(\lambda_\sigma-1)\n\otimes\n)\left(G\bar\nabla c^\pm\mp c^\pm\bar\E\right)\right]. \label{eqn:dimensionless0-4}
\end{align}
In \eqref{eqn:dimensionless0-1}-\eqref{eqn:dimensionless0-4}, the rescaled differential operators are given by
\begin{align}
\bar\nabla=&\left(\eta\partderiv{}{\bar x},  \eta\partderiv{}{\bar y},\partderiv{}{\bar z}\right),\qquad 
\bar\Delta= \eta^2\left(\frac{\partial^2}{\partial \bar x^2} + \frac{\partial^2}{\partial \bar y^2}\right)+ \frac{\partial^2}{\partial \bar z^2}.
\end{align}
Table \ref{param_table} summarizes the parameters of the problem and their experimental values in \cite{Tuxedos}. 

The dimensionless coefficients are
\begin{equation}\begin{array}{cccc}
\displaystyle{B = \frac{\gamma_1T^{-1}}{\ve_0|\ve_a|E_0^2}}, &
\displaystyle{C = \frac{K}{d^2\ve_0|\ve_a|E_0^2}}, &
\displaystyle{ L_1=\frac{e_1-e_3}{\ve_0|\ve_a|E_0d}}, &
\displaystyle{ L_2=\frac{e_1+e_3}{\ve_0|\ve_a|E_0d}},\\
\displaystyle{J = \frac{\ve_\perp}{|\ve_a|}}, & 
\displaystyle{M = \frac{ec_0d}{\ve_0|\ve_a|E_0}}, &
\displaystyle{F = \frac{ K_B\Theta dT^{-1}}{eE_0\bar{D}}=\frac{ec_0d}{E_0\sigma_\perp}}, &
\displaystyle{G = \frac{K_B\Theta}{eE_0d}}.
\end{array} \label{NDconst} 
\end{equation}
Observe that  $J>1$ is always satisfied, guaranteeing the ellipticity of equation (\ref{eqn:dimensionless0-3}) with respect to $\bar\Phi$.
Taking $d=\SI{8}{\micro\meter}$,  postponing the choice of  {$T$}, and using the remaining values in table \ref{table-parameter}, we have
\be \begin{array}{llll}
B = (\SI{2.4e-5}{\second}) T^{-1}, &
C = 6.25\times 10^{-5}, &
J = 2.1, &
M = .84,\\
F = (\SI{5.1e-3}{\second})T^{-1}, &
	G = 4.1\times 10^{-4}, &
	\lambda_\sigma = 0.8, & \eta = 1.6\times10^{-3}. \end{array} \label{constant-values}
\ee
The flexoelectric constants $e_1$ and $e_3$ are difficult to measure experimentally, but they are on the order of $\SI{e-11}{\coulomb\per\meter}$ \cite{castles2012flexoelectric}, \cite{cheung2004calculation}. Therefore the flexoelectric coefficients obey $| L_1|,| L_2|\approx L_f$, where
\be
L_f := \frac{\SI{e-11}{\coulomb\per\meter}}{\ve_0\ve_aE_0d} = 4.1\times 10^{-3}. \label{Lf}
\ee
\noindent Prior to choosing the value of $T$ in the dimensionless parameter groups \eqref{constant-values}, we determine the time scales associated with the different effects of the model. 
\begin{enumerate}
\item Letting $\omega$ be the frequency of the applied electric field, the corresponding time constant  $T_\omega=\frac{2\pi}{\omega}$  falls within the range
\begin{equation}
 0.50\pi\times 10^{-2} \text{s}\geq T_\omega\geq 0.44\pi \times 10^{-2} \text{s}. \label{frequency-range} \end{equation}
\item The scale of dielectric effects follows from the relation $B=1$ and gives $T_{\text{dielc}}=\SI{0.21e-4}{\second}$.
\item The scale of the flexoelectric effect follows from equating $B=L_f$, giving $T_{\text{flex}}=\SI{0.59e-2}{\second}$.
\item The scale of the elastic effect follows from setting $B=C$ and gives
$T_{\text{elast}}=\SI{0.38}{\second}$.
\end{enumerate}
The previous calculations indicate that, within the middle to high  frequency range,  the applied electric field activates the flexoelectric affects with a time scale of its own order of magnitude. This motivates us to choosing
\be
T= 2\pi\omega^{-1}. \label{Tomega}
\ee
We also find that the dielectric effects relax   faster,  whereas the elastic ones evolve within a greater  time scale. 
 Typical values of the dimensionless constants, that will be taken as reference in the analysis, are calculated with $\omega$ as in (\ref{frequency-range}):
\be\centering\begin{aligned}
&B = 1.2\times 10^{-2},\qquad
C = 6.3\times 10^{-5},\qquad
J = 2.1,\qquad
M = 0.84,\\
& F = 2.6,\qquad
G = 4.1\times 10^{-4},\qquad
| L_1|,| L_2|\approx 4.1\times 10^{-3},\qquad
\lambda_\sigma = 0.8.\label{coefficients-numerical}
\end{aligned} \ee
The scaled equations also reveal the  dielectric effect as dominant and, with the flexoelectric one being between $10^2$ and $10^3$ times smaller. Moreover, for a sufficiently large electric field strength $E_0$, 
$| L_1|$ and $| L_2|$ become very small, suppressing the flexoelectric mechanism. At such a limit, the appearance of convection would be expected. 
Next, we investigate the speed $s$ of the soliton-like structure. In \cite{Tuxedos}, the authors report speeds in the range of 150 to 400 $\mu$m/sec. With the length of the electrode plate taken as $5.5$ mm, the range of time scales $T_s$ of the soliton motion is 
\begin{equation}
    T^s_\text{min}=\frac{5.5}{0.4}=14 \text{s} \leq T_s\leq  T^s_\text{max}=\frac{5.5}{0.1}=55 \text{s}.  \label{solition-time-scale}
\end{equation}
The dimensionless form of the soliton speed is taken as 
\be
R = \frac{s}{L\omega}=\frac{\eta s}{\omega d}. \label{dimensionless-soliton-speed}
\ee 
Within the observed speed interval and the frequency range of the applied electric field, we find that 
\be
R_\text{min}=\frac{100}{5\times10^3\times 450}=0.444\times 10^{-4}\leq R\leq R_\text{max}=\frac{450}{5\times 10^3\times 400}=2.25\times 10^{-4}. \label{dimensionless-soliton-speed-range} \ee
It remains to specify the initial and boundary conditions of the problem \eqref{eqn:dimensionless0-1}-\eqref{eqn:dimensionless0-4}. These are
\be
\theta(x,y,z, 0)=\theta_0(\bar{x},\bar{y},\bar{z}),\qquad
\phi(x, y, z, 0)=\phi_0(\bar{x},\bar{y},\bar{z}),\qquad
\bar{c}^\pm(x, y, z, 0)=\bar{c}^\pm_0(\bar{x},\bar{y},\bar{z}). \label{244}
\ee
At $\bar x=\pm\tfrac{1}{2}$, $\bar{y}=0,1$, and $\bar{z}=0,1$,
\be
\theta=\phi=0,\qquad \bar{c}^+-\bar{c}^-=0,\qquad \bar{c}^++\bar{c}^-=Q_0 \qquad\text{for all }\bar{t}\geq 0.
\ee
Additionally,

\begin{gather}
\bar{\Phi}(\bar x, \bar y, 1, \bar t )=\bar{\Phi}_0(\bar{t}),\qquad \bar{\Phi}\bar x, \bar y, 0, \bar t)=0, \qquad\bar x, \bar y\in \bar\Omega_\perp\\
\bar\Phi_{\boldsymbol\nu}\left(\bar x, \bar y, \bar z, \bar t\right)=0, 
\quad \bar x, \bar y\in \partial\bar\Omega_\perp, \quad 0<\bar z<1, \label{246}\end{gather}
where $\boldsymbol{\nu}$ denotes the unit outer normal to the boundary $\partial\bar{\Omega}_\perp$ (where corners are being excluded).
The constant $Q_0$ denotes a background charge concentration (both signs) representing the amount of impurities in the system. Assuming that there are two ion species  in the sample, we  take $Q_0=2$. Dirichlet boundary conditions on $\n$ express the strong anchoring imposed on the bounding plates. Likewise, prescribing the electric potential on the plates is compatible with the waveform generator used in the experiment. For the charges, the assumption of Dirichlet boundary conditions instead of the standard no-flux is done for analysis convenience. Note that the initial profile of $\bar{\Phi}$ can be computed from Poisson's equation and the conditions \eqref{244}-\eqref{246}.

\subsection{Traveling wave geometry, symmetry and time multiscale}
We study traveling wave solutions that move along the $y$-axis  with positive dimensionless speed $R>0$ as in (\ref{dimensionless-soliton-speed}). We therefore define the similarity variable
\be
\bar\xi = \bar y-R\bar t,
\ee
However, for our current analysis, we treat $R$ as one of the unknowns of the problem. 

From this point forward,  we suppress the superimposed bar notation on variables and look for solutions that depend on the variables $t$, $x$, $z$, and $\xi$. The partial derivatives transform as
\be
\frac{\partial}{\partial y} = \frac{\partial\xi}{\partial y}\frac{\partial}{\partial \xi} = \frac{\partial}{\partial \xi},\qquad
\frac{d}{dt} = \frac{\partial}{\partial t}+\frac{\partial\xi}{\partial t}\frac{\partial}{\partial\xi}=\frac{\partial}{\partial t}-R\frac{\partial}{\partial\xi},
\ee
and the corresponding gradient and Laplacian operators are
\be
\nabla_\xi=\left(\eta\frac{\partial}{\partial x}, \eta\frac{\partial}{\partial \xi}, \frac{\partial}{\partial z}\right), \qquad \Delta_\xi= \eta^2\left(\frac{\partial^2}{\partial x^2}+ \frac{\partial^2}{\partial \xi^2}\right) + \frac{\partial^2}{\partial z^2}.
	\ee
The governing system  (\ref{eqn:dimensionless0-1})-(\ref{eqn:dimensionless0-4}) becomes
\be\left\{\begin{aligned}
&B\phi_t
=  BR\phi_\xi+C\left(\Delta_\xi\phi-2\tan\theta\;\nabla_\xi\phi\cdot\nabla_\xi\theta\right)
+(\n\cdot\E)(\E\cdot\m)\sec\theta+ L_2[\m\cdot(\nabla_\xi\E)\n]\sec\theta \\
&\qquad\qquad\qquad
- L_1(\nabla_\xi\theta\cdot\m)(\E\cdot\p)\sec\theta
+ L_1(\nabla_\xi\theta\cdot\p)(\E\cdot\m)\sec\theta
, \\
&  B\theta_t
=  BR\theta_\xi+C\left(\Delta_\xi\theta+\tfrac{1}{2}\sin2\theta\;|\nabla_\xi\phi|^2\right)
+(\n\cdot\E)(\E\cdot\p)+ L_2[\p\cdot(\nabla_\xi\E)\n]\\
&\qquad\qquad\qquad
- L_1(\nabla_\xi\phi\cdot\p)(\E\cdot\m)\cos\theta
+ L_1\cos\theta(\nabla_\xi\phi\cdot\m)(\E\cdot\p)
,\\
& \nabla_\xi\cdot\left[J\E-(\n\cdot\E)\n+\tfrac{1}{2}( L_1+ L_2)(\nabla_\xi\cdot\n)\n+\tfrac{1}{2}( L_2- L_1)(\nabla_\xi\n)\n\right]=Mq, \\
& F Q_t = FRQ_\xi+\nabla_\xi\cdot\left[\left(\mathbf{I}+(\lambda_\sigma-1)\n\otimes\n\right)\left(G\nabla_\xi Q- q\E\right)\right],\\
& F q_t =FRq_\xi+\nabla_\xi\cdot\left[\left(\mathbf{I}+(\lambda_\sigma-1)\n\otimes\n\right)\left(G\nabla_\xi q- Q\E\right)\right],
\end{aligned}\right. \label{NDsystem} \ee
where  the dimensionless variables \be Q := c^++c^-,\qquad  q:=c^+-c^- \ee
denote the total unsigned background charge and the net charge, respectively. 
The unknown dimensionless fields of the problem are $\phi$, $\theta$, $\Phi$, $Q$, and $q$.

We point out that choosing $R>0$ implies the selection of a disturbance moving in the positive $y$-direction. However, the experiments show that there are also chevrons moving in the opposite direction. Indeed, a simple calculation shows that, if there exists a chevron moving along the positive $y$-direction, there is a symmetric one moving opposite to it. We formulate this feature as follows:

\subsubsection{Symmetry property of the solutions.} Suppose $(\phi,\theta,\Phi,q)$ is a solution to \eqref{NDsystem} with speed $R$. Then
\be\left\{\begin{aligned}
	\phi^*(x,\xi,t) &= -\phi(x,-\xi,t),\quad 
	\theta^*(x,\xi,t) = \theta(x,-\xi,t),\\
	&\Phi^*(x,\xi,t) = \Phi(x,-\xi,t),\\
	q^*(x,\xi,t) &= q(x,-\xi,t),\quad 
		Q^*(x,\xi,t) = Q(x,-\xi,t).
\end{aligned}\right.\ee
is a solution to \eqref{NDsystem} with speed $-R$. This property establishes that, for every soliton-like package moving with speed $R$ along the positive direction, there is another one, with the opposite {\it bullet}
profile (Figure \ref{fig1}, top right illustration) that moves with the same speed along the negative direction. 

\subsubsection{Initial layer property}
We conclude this section observing that the governing system (\ref{NDsystem}) has 
two main time scales,  relevant to the dynamics of the angular profile $\phi$ and $\theta$. These are the standard dimensionless time $t$ and the {\it fast } time  $\hat t:=\frac{t}{B}$, with $B$ as in (\ref{constant-values}). This motivates us to consider solutions of the system such that $\theta=\theta(t, \hat t, x, \xi, z) $ and $\phi=\phi(t, \hat t, x, \xi, z)$. Furthermore, the size of the coefficients of the system, together with standard asymptotic arguments associated to initial layer analysis, allow us to approximate the angular equations of the governing system in the $\hat t$ scale as
\begin{align} \phi_{\hat t} =&(\n\cdot\mathbf E)(\mathbf E\cdot \m)\sec\theta, \quad 
 \theta_{\hat t}= (\n\cdot\mathbf E)(\mathbf E\cdot \p), \quad \hat t:=\frac{t}{B}. \label{initial-layer0}
 \end{align}
 In particular, this indicates that,  the initial conditions on the shape of the distortion that triggers the chevron are only felt at the very early time stages of the process, near  $t=0$,  (lasting about $10^{-2}$, in dimensionless time). However, due to the periodicity of the source potential, the effect also reappears at $t= n\pi$, $n\geq 0$, integer, where $|p(n\pi)|=|\cos(n\pi)|=1$. 
 This indicates that the dielectric effects,  which depend only on the size of the applied field, are present in the system, discretely in time, manifesting themselves in a periodic fashion, with their action lasting about $10^{-2}$ seconds. On the other hand, the soliton disturbance is almost  entirely shaped by flexoelectric, viscous and elastic effects. 
 We will revisit this property in reference to the linear system.

\section{Instability of the uniform state} We take the point of view that soliton disturbances emerge at the unstability onset of the uniform state. For this, we perform a stability analysis of such a state to determine the instability threshold, and the corresponding lengths and time scales associated with it. For this, we linearize the governing system 
 \eqref{NDsystem}
about the uniform state
\be
\phi=\theta=0, \qquad \Phi = p(t)z,\qquad q=0,\qquad Q=Q_0, \label{linearBase}
\ee
where
\be \E=-\nabla \Phi \quad \text {and}\quad 
p(t)=\cos{t}. \label{linearBase2}
\ee
Note that the solution \eqref{linearBase}-\eqref{linearBase2} satisfies the initial and boundary conditions \eqref{244}-\eqref{246} with $\theta_0=\phi_0=0$, $c^\pm_0=\tfrac{1}{2}Q_0$, and $\Phi_0=p(t)$. This choice of $\Phi$ corresponds to an alternating electric field in the $z$-direction, as in the experiments in \cite{Tuxedos}. Our goal is  to show that this uniform solution is unstable to chevron-like traveling waves. For this, we proceed in several steps, that include approximating the original system as follows: 
\begin{enumerate}
    \item Linearize the system about the equilibrium solution.
    \item Since the domain aspect ratio $\eta$ is of the order of $10^{-3}$, we average the previously obtained system along the direction perpendicular to the plates.
    \item We take the Fourier transform of the resulting system with respect to the space variables $x$ and $\xi$.
    \item We formulate the conditions on the combined parameters and Fourier modes that lead to neutral stability, and analyze the resulting relations. 
\end{enumerate}
The linear system, with respect to the time scale $t$ is
\be\left\{\begin{aligned}
& B\phi_t = BR\phi_\xi+ C\bar\Delta_\xi\phi{{+}} L_1\eta p(t)\theta_\xi{+} L_2\eta^2\Phi_{x\xi},\\
& B\theta_t =  BR\theta_\xi+C\bar\Delta_\xi\theta-p(t)\eta\Phi_x- L_1\eta p(t)\phi_\xi{+} L_2\eta^2\Phi_{xz},\\
& (J-1)\eta^2\Phi_{xx}+J\eta^2\Phi_{\xi\xi}+J\Phi_{zz} = p(t)\eta\theta_x+ L_2\eta^2\phi_{x\xi}+\eta\theta_{xz}-Mq,\\
& Fq_t =  FR q_\xi+G(\lambda_\sigma\eta^2 q_{xx}+\eta^2q_{\xi\xi}+q_{zz})+p(t)Q_z\\
&\textcolor{white}{fq_t}\qquad +Q_0\left[\lambda_\sigma\eta^2\Phi_{xx}+\eta^2 \Phi_{\xi\xi}+\Phi_{zz}+(\lambda_\sigma-1)p(t)\eta\theta_x\right],\\
& FQ_t = FRQ_\xi+G(\lambda_\sigma\eta^2 Q_{xx}+\eta^2Q_{\xi\xi}+Q_{zz})+p(t)q_z. 
\end{aligned}\right.\label{Linearized}\ee
\begin{remark}
Observe that if the flexoelectric terms are removed ($ L_1= L_2=0$), the equation for $\phi$ is simply
\be
B\phi_t=BR\phi_\xi + C\Delta_\xi \phi,
\ee
which is the heat equation with a lower-order terms. Without flexoelectricity, there is no forcing in the equation, and the $\phi$ profile will dissipate to zero. However, the experiments show that the $\phi$ profile is roughly constant in time. This provides strong evidence that the flexoelectricity is indeed responsible for the formation of the chevrons.
\end{remark}
\subsection {Space averaging}
The averaging performed next, while reducing the problem to the two-dimensional space variables $x$ and $\xi$ is consistent with the three-dimensional nature of the chevron  \cite{Tuxedos}. In particular, it   allows for the trivial boundary conditions on the angular and concentration variables to hold on the boundary plates $z=0$ and $z=1$, and the electric field  taking the prescribed values on the electrodes.  This, in turn, is consistent with the observation that the director disruption occurs in a thin region around the middle of the domain. 
Motivated by the techniques of  approximating functions by sums of orthogonal polynomials,  we assume that the  $z$-dependence of the fields follows  the parabolic profile,
\be\begin{aligned}\label{eqn:perturbations}
& \phi=r(z)\tilde{\phi}(x,\xi,t),\qquad
\theta=r(z)\tilde{\theta}(x,\xi,t),\qquad
\Phi = r(z)\tilde{\Phi}(x,\xi,t),\\
& q=r(z)\tilde{q}(x,\xi,t),\qquad
Q=r(z)\tilde{Q}(x,\xi,t),
\end{aligned}\ee
where $r(z) = 6z(1-z)$. Then
\be
\fint_0^1 r(z)\;dz=1,\qquad
\fint_0^1 r'(z)\;dz=0,\qquad
\fint_0^1 r''(z)\;dz=-12,
\ee
so after averaging in $z$ (and dropping the tildes), the linear system \eqref{Linearized} becomes
\be\left\{\begin{aligned}
& B\phi_t = BR\phi_\xi+C(\eta^2\phi_{xx}+\eta^2\phi_{\xi\xi}-12\phi)
{+} L_1\eta p(t)\theta_\xi{+}
 L_2\eta^2\Phi_{x\xi},\\
&B\theta_t =  BR\theta_\xi+C(\eta^2\theta_{xx}+\eta^2\theta_{\xi\xi}-12\theta)-p^2(t)\theta
-p(t)\eta\Phi_x- L_1\eta p(t)\phi_\xi,\\
& (J-1)\eta^2\Phi_{xx}+J\eta^2\Phi_{\xi\xi}-12J\Phi = p(t)\eta\theta_x+\eta^2  L_2\phi_{x\xi}-Mq,\\
& Fq_t =  FRq_\xi+ G(\lambda_\sigma \eta^2q_{xx}+\eta^2q_{\xi\xi}-12q)
+Q_0(\lambda_\sigma\eta^2\Phi_{xx}+\eta^2\Phi_{\xi\xi}-12\Phi)+Q_0(\lambda_\sigma-1)p(t)\eta\theta_x.
\end{aligned}\right. \label{zsys}\ee
and
\be
F{Q}_t=FR{Q}_\xi + G(\lambda_\sigma\eta^2 {Q}_{xx}+\eta^2{Q}_{\xi\xi}-12{Q}) \label{Qhat}
\ee
The equation \eqref{Qhat} for $Q$ decouples from the rest of the system, so it can be disregarded. The total concentration of ions enters \eqref{zsys} only through the constant $Q_0$.

\subsection{Fourier analysis}
We adopt the method of the Fourier transform to determine instability thresholds of the base solution (\ref{linearBase})-(\ref{linearBase2}). The perturbation functions are twice continuously differentiable and are also elements of $L^1(\mathbb R^2)$. 
This assumptions are consistent with two of the main aspects of the phenomenon: the triggering body force mechanism due to the applied electric field, rather than  conditions at the boundary, and the persistence of the uniform base configuration away from the plane center region. This justifies the   trivial extension of the
base solution to the entire $x-\xi$ plane.

We start with  applying the Fourier transform to \eqref{zsys} in both $x$ and $\xi$, i.e.
\be
\hat{f}(\rho_x,\rho_\xi) = \int_{\rr^2}f(x,y)\text{ exp }\left(-2\pi i(\rho_xx+\rho_\xi\xi)\right)\;dx\;d\xi.
\ee
Poisson's equation gives
\be
\hat{\Phi} = \frac{1}{4\Delta_J(\rho_x,\rho_\xi)}\left(4\pi^2\eta^2  L_2\rho_x\rho_\xi\hat{\phi}-2\pi i \eta p(t)\rho_x\hat{\theta}+M\hat{q}\right) \label{FTGauss}
\ee
with
\be
\Delta_J(\rho_x,\rho_\xi) = (J-1)\eta^2\pi^2\rho_x^2+J\eta^2\pi^2\rho_\xi^2+3J. \label{Delta_J}
\ee
In order to solve \eqref{FTGauss}, we assume that $\Delta_J(\rho_x,\rho_\xi)\neq 0$. Substituting \eqref{FTGauss} into the  remaining equations for $\hat{\phi}$, $\hat{\theta}$  and $\hat{q}$ in (\ref{zsys}) yields
\begin{align}
B\hat{\phi}_t &= \left[2\pi i BR\rho_\xi-4C\Delta_1(\rho_x,\rho_\xi)-\frac{4\pi^4 L_2^2\eta^4\rho_x^2\rho_\xi^2}{\Delta_J(\rho_x,\rho_\xi)}\right]\hat{\phi}\nonumber\\
&\hspace{2in}
+2\pi i\eta\rho_\xi p(t)\left(  L_1+\frac{\pi^2 L_2\eta^2\rho_x^2}{\Delta_J(\rho_x,\rho_\xi)}\right)\hat{\theta}
-\frac{\pi^2 L_2M\eta^2\rho_x\rho_\xi}{\Delta_J(\rho_x,\rho_\xi)}\hat{q},\label{FourierA}\\
B\hat{\theta}_t &= \left[2\pi i BR\rho_\xi-4C\Delta_1(\rho_x,\rho_\xi)-\left(1+\frac{\pi^2\eta^2\rho_x^2}{\Delta_J(\rho_x,\rho_\xi)}\right)p^2(t)\right]\hat{\theta} \nonumber\\
&\hspace{2in} -2\pi i \eta\rho_\xi\left( L_1+\frac{ L_2\pi^2\eta^2\rho_x^2}{\Delta_J(\rho_x,\rho_\xi)}\right)\hat{\phi}p(t)
-\frac{\pi i M\eta\rho_x}{2\Delta_J(\rho_x,\rho_\xi)}\hat{q}p(t)
\label{Fourier}\\
F\hat{q}_t &= \left(2\pi i FR\rho_\xi-4G \Delta_\sigma(\rho_x,\rho_\xi)-\frac{Q_0M\Delta_\sigma(\rho_x,\rho_\xi)}{\Delta_J(\rho_x,\rho_\xi)}\right)\hat{q}
-4\pi^2\eta^2Q_0 L_2\rho_x\rho_\xi\frac{\Delta_\sigma(\rho_x,\rho_\xi)}{\Delta_J(\rho_x,\rho_\xi)}\hat{\phi}\nonumber\\
&\hspace{2in}
+2\pi i\eta Q_0\rho_x\left(\lambda_\sigma-1+\frac{\Delta_\sigma(\rho_x,\rho_\xi)}{\Delta_J(\rho_x,\rho_\xi)}\right)\hat{\theta}p(t),\label{FourierB}
\end{align}
where
\be
\Delta_\sigma(\rho_x,\rho_\xi) = 3+\pi^2\eta^2(\lambda_\sigma\rho_x^2+\rho_\xi^2),\qquad
\Delta_1(\rho_x,\rho_\sigma) = 3+\pi^2\eta^2(\rho_x^2+\rho_\xi^2). 
\ee
Our goal is to identify neutrally stable solutions of the previous system 
that also allow for the sustained time oscillation of the out of plane angular variable $\theta$.
Now, let us formally write the system as 
\begin{equation}\u_t=\A(t)\u,\qquad\qquad
\u=\begin{bmatrix} \hat{\phi} & \hat{\theta} & \hat{q}\end{bmatrix}^\intercal, \label{eqn:linear-system1}
\end{equation}
where $\A$ is directly obtained from the coefficients of equations (\ref{FourierA})-(\ref{FourierB}). Suppressing the dependence of $\Delta_1$, $\Delta_J$, and $\Delta_\sigma$ on $\rho_x$ and $\rho_\xi$, we find that the components of the matrix $\A$ are:
\begin{align} 
A_{11} &= V-a_{11},  \quad A_{12}(t) = -A_{21}(t)
=\frac{2\pi i\eta\rho_\xi}{B\Delta_J}\left[JL_1\Delta_1+\pi^2\eta^2(L_2-L_1)\rho_x^2\right]p(t), \nonumber  \\ 
A_{13}  = &-\frac{\pi^2 L_2M\eta^2\rho_x\rho_\xi}{B\Delta_J}, \quad  A_{22}(t)
= V-a_{22}(t),                  \quad                     A_{23} = {-\frac{\pi i M{\rho}_x\eta}{2B\Delta_J}p(t)},  \nonumber \\
 A_{31} =& -\frac{4\pi^2\eta^2Q_0L_2\rho_x\rho_\xi}{F}\frac{\Delta_\sigma}{\Delta_J}, \quad 
 A_{32}(t) = \frac{2\pi i\lambda_\sigma Q_0\eta\rho_x}{F}\frac{\Delta_1}{\Delta_J}p(t), \quad 
 A_{33} = V-a_{33},  \label{Aij}
\end{align}
where 
\begin{align}
V=&  2\pi i R\rho_\xi, \quad   a_{11}=\frac{4}{B}\left(C\Delta_1+\frac{\pi^4 L_2^2\eta^4\rho_x^2\rho_\xi^2}{{\Delta_J}}\right),\label{V}\\
 \quad a_{22}(t)=&\frac{4C}{B}\Delta_1-\frac{J}{B}\frac{\Delta_1}{\Delta_J}p^2(t):=\tilde a_{22}-\frac{J}{B}\frac{\Delta_1}{\Delta_J}p^2(t), \quad a_{33}= 
\frac{\Delta_\sigma}{F}\left(4G+\frac{Q_0M}{\Delta_J}\right). \label{aij}
\end{align}
\begin{remark} \label{initial-layer} We point out that the second term in $a_{22}(t)$ in \eqref{aij} is of order $O(1)$, except for the special wave numbers on the line $\Delta_1=0$. Away from the latter case, the separation of time scales, between $t$ and $\frac{t}{B}$, indicates that the $p^2(t)-$term  contributes to the initial layer of the solution $\theta,$ and as such, it should be  treated separately from the time-$t$ dynamics, as customary in initial layer analyses \cite{o1991singular}.  Consequently, from now on, we will omit 
the $p^2(t)-$term from $A_{22}$ and $a_{22}$ and replace them with $\tilde A_{22}$ and $\tilde a_{22}$, respectively. \end{remark} 

\noindent In order to analyze the stability of the system \eqref{eqn:linear-system1}-\eqref{Aij}, we restrict the wave number variables to purely imaginary, that is, 
\be
\text{Re }\rho_x=\text{Re }\rho_\xi=0. \label{assume}
\ee
The condition \eqref{assume} simplifies the calculations, but it also matches the experimental findings. Since the chevrons do not exhibit periodic behavior in either the $x$- or $y$-directions, we expect the real parts of the Fourier variables $\rho_x$ and $\rho_\xi$ to be zero. Taking $\rho_x$ and $\rho_\xi$ purely imaginary means that solutions display exponential growth or decay in both the $x$- and $y$-directions.
We note  that, under the assumption \eqref{assume}, all the components of $\A$ in \eqref{Aij} are real-valued.
Let us introduce the notation 
\be
c_x = -\pi^2\eta^2\rho_x^2,\qquad c_\xi=-\pi^2\eta^2\rho_\xi^2 \label{cx-cxi}
\ee
If follows from the condition (\ref{assume}) that \, $c_x, c_\xi\geq 0$. 
\smallskip
The choice of the sign of $\text{Im}\rho_\xi $ will be done later, when sorting out the sign of $R$.
%
%

\begin{proposition} Let us consider perturbations satisfying (\ref{assume}). Then, 
the equilibrium solution \eqref{linearBase}-\eqref{linearBase2} of the problem (\ref{eqn:linear-system1}), for parameter data as in (\ref{coefficients-numerical}), is unstable for perturbations whose speed and wave numbers satisfy 
\be\begin{aligned}
 -6\pi  R\,\text{Im}\,\rho_\xi-& H(c_x, c_\xi)\geq 0 \quad \text{and} \quad \Delta_J\neq 0, \\
 H(c_x, c_\xi):= &\frac{1}{B\Delta_J}[8C\Delta_1\Delta_J+ 4L_2^2c_xc_\xi\Delta_J +\frac{J}{2}\Delta_1 +\frac{\Delta_\sigma}{F}(4G+BQ_0M)],\end{aligned} \label{324} \ee

and, with $c_x$ and $c_\xi$ as in (\ref{cx-cxi}). 
\end{proposition}
\begin{proof}
Let $\mu_1$, $\mu_2$, and $\mu_3$ denote the Floquet exponents of the system \cite{Hale}. Recall the trace property  of the Floquet theory of linear systems with periodic coefficients, that is,
\begin{align}
\sum_{i=1}^3\mu_i = &\fint_0^1\text{tr }\mathbf{A}(t)\;dt,  \label{traceFormula}\\
\tr(\A)=& 6\pi i R\rho_\xi-\frac{8C\Delta_1(\rho_x,\rho_\xi)}{B}
-\frac{4\pi^4L_2^2\eta^2\rho_x^2\rho_\xi^2+Jp^2(t)}{B\Delta_J(\rho_x,\rho_\xi)}
-\frac{\Delta_\sigma(\rho_x,\rho_\xi)}{F}\left(4G+\frac{Q_0M}{\Delta_J(\rho_x,\rho_\xi)}\right). \label{traceA}
\end{align}
The result follows by integrating \eqref{traceA} with respect to $t$ and taking into account that the diagonal elements are all constant, except for the $p^2(t)$ term appearing in the component $A_{22}$ of \eqref{Aij}, and then setting $\fint_0^1\text{tr }\A(t)~dt\geq 0$. The positivity of \eqref{324} ensures that at least one Floquet exponent has positive real part.
\end{proof}
\begin{remark} Inequality \eqref{324} provides a useful insight in finding the threshold conditions that trigger the instability of the uniform solution. For this, we need to identify   wave number  parameters that satisfy the equation $H(c_x, c_\xi)=0$, for perturbations with speed $R=0$. In a later section, we will further interpret the wave number equation in terms of  the eigenvalues of the system that characterize the neutral stability of the uniform solution. 
\end{remark}


We summarize the main results of this work on the following theorem. For this, we first rewrite  the linear system \eqref{FourierB} in the form
\be
\dot\u= \A^0\u +\mathbf h(t), \label{LS-main}
\ee 
 where 
 \begin{align}
 A^0_{ij}= &A_{ij}, \,   i=j=1, 3; \, \,  A^0_{22}=\tilde A_{22},  \nonumber\\  A^0_{13}= & A_{13}, \, \, A^0_{23}= A_{23}, \,\, A^0_{31}=A_{31}, \, 
 A^0_{12}=0, \, A^0_{21}=0, \nonumber\\
 \mathbf h=&\left[\begin{matrix} A_{12}(t)\hat\theta & A_{21}(t)\hat\phi & A_{32}(t)\hat\theta \end{matrix}\right]^T,
 \end{align}

\begin{theorem}  Suppose that $\epsilon_a<0$ and $\lambda_\sigma<1$, and that  the parameters of the problem are as in (\ref{constant-values}).  There exist wave numbers $\rho_x, \rho_\xi$ and perturbation speeds $R$, for which $\lambda_1\geq \lambda_3\geq 0$, where $\lambda_i, i=1,3, $ denote the eigenvalues of the matrix $A^0$, so that the solution $\u=\left[\begin{matrix} 0 &0& 0\end{matrix}\right]^T$ of the Fourier system \eqref{FourierB}   is an unstable  node.  Moreover, for the wave numbers and perturbation speeds satisfying the stricter relations  \eqref{threshold-conditions},  $\lambda_1\approx 0\approx \lambda_3$ hold up to terms of the order $O(10^{-5})$. 
Furthermore, the perturbations that solve the approximate linear system obtained from equation \eqref{FourierB} by time averaging of two appropriate off-diagonal terms have the property that the out-of plane angle $\hat\theta $ has a nonzero phase-shift of approximately $\frac{\pi}{2}$ radians with respect to the applied electric field. 
Moreover, the size of the unstable perturbations and the chevron speed fall within the experimental range.  
\end{theorem}

\begin{remark}
	The requirement that \eqref{threshold-conditions} hold ensures the  vanishing of the eigenvalues $\lambda_1$ and $\lambda_3$. Equivalently, it amounts to a selection of  wave numbers	and speeds of the perturbations in the neutral stability regime, that is, at the instability threshold, that preserve the shapes of the components $\phi$ and $q$.
\end{remark}

The proof is carried out in the next subsection, and it proceeds in four steps. 
For notational convenience, we will suppress the 'hat' symbol from the fields $\phi, \theta, q.$
\subsection {Solution of the linear system}
\noindent
Next, we solve the system \eqref {eqn:linear-system1} with respect to the time variable $t$, with the wave numbers playing the role of parameters of the system. We further identify ranges of such parameters that lead to neutral stability of the solutions.
Two approximations will enter the analysis, one involving time averaging of two off diagonal terms of the order  $O(L_i)$. 
A second one, that replaces the positive term $p^2(t)$, on the main diagonal of the system,  by its one-period average $\frac{1}{2}$, is done in order to replace an otherwise longer calculation.
 
\noindent{\bf Step 1.\,} We start solving the homogeneous system \eqref{LS-main}. 
It corresponds to the three ordinary differential equations
\begin{align}
    \left[\begin{matrix} \phi_t \\ q_t\end{matrix}\right] =   &\left[\begin{matrix} A_{11}&A_{13}\\ A_{31}&A_{33} \end{matrix}\right]  \left[\begin{matrix} \phi \\ q\end{matrix}\right], \label{2by2}\\
    \theta_t=&A_{22}^0\theta + A_{23}(t)q. \label{theta-q}
\end{align}
The characteristic equation associated with the problem \eqref{2by2} is
\be \lambda^2-\tr(A^0) \lambda+\det A^0=0.\ee 
The roots are 
\be \lambda_1=\frac{1}{2}[\tr(A^0) + \sqrt{\tr^2(A^0)-4\det A_0}], \quad  \lambda_3=\frac{1}{2}[\tr(A^0) - \sqrt{\tr^2(A^0)-4\det A_0}]. \label{e-values} \ee
Let us rewrite
\be \lambda_{1,3}=\frac{1}{2}[(A_{11}+A_{33})\pm \sqrt{{(A_{11}-A_{33})}^2+ A_{13}A_{31}}]. \label{e-values-reduced}\ee
The corresponding eigenvectors are
\be \v_1=\left[\begin{matrix} \omega_1\\1\end{matrix}\right], \quad  \v_3=\left[\begin{matrix} \omega_3\\1\end{matrix}\right], \ee
where 
\be \omega_1:= -\frac{A_{13}}{A_{11}-\lambda_1}, \quad 
\omega_3:= -\frac{A_{13}}{A_{11}-\lambda_3}, \quad \omega_1\neq \omega_3. \ee
The general solution to the system \eqref{2by2} is 
\be \left[\begin{matrix}\phi(t)\\ q(t)\end{matrix}\right]= a_0 e^{\lambda_1 t}\v_1+ b_0e^{\lambda_3 t}\v_3, \label{phi-q} \ee
where $a_0$ and $b_0$ are arbitrary constants. 
Note that 
two relevant cases arise according to the nature of the eigenvalues $\lambda_1, \lambda_3$:
\begin{align}
   {\text(i)}:\,  {\tr^2(A^0)-4\det A_0}\geq 0, \quad \text{and}\quad  {\text(ii)}: \,{\tr^2(A^0)-4\det A_0}< 0. \label{discrim}
\end{align}
They correspond to the equilibrium solution  $[0,0]^T$ of \eqref{2by2} being,  (i) a node or saddle point,  or (ii) a spiral. 
To help us determine the wave number ranges consistent with the soliton-lik instability, we set up the following lemma.
\begin{lemma}\label{lemma3.1} Assume that $\tr \A_0\geq 0. $ Let us consider wave numbers such that (\ref{assume}) holds. 
Then for both cases  (i) and (ii) in \eqref{discrim} the   $[0,0]^T$ solution of \eqref{2by2} is unstable.  Furthermore, for parameter values as in \eqref{coefficients-numerical}, the product
\be A_{13}A_{31}= \frac{L_2^2MQ_0}{BF}\frac{\Delta_\sigma}{\Delta_J^2} c_xc_\xi= 0\, \Longleftrightarrow \, \Delta_\sigma= 0. \label{positive-discrim} \ee
\end{lemma}
Let us consider the wave number range  such that \be \lambda_1\geq\lambda_3\geq 0 \label{unstable-node}. \ee
We will find that such a range is consistent with \eqref{positive-discrim} being satisfied to the order $O(L_i^2\times 10^{-2})$.  In such a case, $[0,0]^T$ is an unstable node. Let us now consider the solution to \eqref{2by2} along the invariant line $\v_1=0, $ and apply it to equation \eqref{theta-q}, that now takes the form
 \be\theta_t=A_{22}^0\theta +b_0 A_{23}(t)  e^{\lambda_3 t}. \label{soln-2by2-general}\ee 
 Note that departure from the branch $\v_1$ eliminates the highest decay rate, $\lambda_1$, from the solution. 
The general solution of the latter equation  is then
\begin{align}\Theta(t)=&\exp{(\int_0^tA_{22}^0(s)\,ds)}\big(C_0+ X(t)), \\
X(t):=& b_0\int_0^t\exp[-\int_0^s A_{22}^0(u)\,du]\,A_{23}(s) e^{\lambda_3s} \,ds, \end{align}
where $b_0$ and $C_0$ are arbitrary constants. 
 We also make the simplifying assumption of replacing 
$A_{22}(t)$ by its average on the interval $[0,1]$, and we arrive at 
\be \begin{aligned} X(t)=&\int_0^te^{(\lambda_3-A_{22}^0)s}A_{23}(s)\,ds\nonumber \\ =&\frac{M\sqrt{c_x}}{2B\Delta_J}\frac{e^{(\lambda_3-A_{22}^0)t}}{(\lambda_3-A_{22}^0)^2+4\pi^2}[2\pi\sin{2\pi t}+ (\lambda_3-A_{22}^0)\cos{2\pi t}]. \end{aligned}\ee
Hence, 
\be \Theta(t)= C_0e^{A_{22}^0t}+ b_0\frac{M\sqrt{c_x}}{2B\Delta_J}\frac{e^{\lambda_3t}}{(\lambda_3-A_{22}^0)^2+4\pi^2}[2\pi\sin{2\pi t}+ (\lambda_3-A_{22}^0)\cos{2\pi t}]. \label{Theta-final}\ee
We immediately observe the phase shift of the out-of-plane angular component with respect to the applied AC field, that we shall estimate later.

A fundamental matrix solution of the system \eqref{LS-main} and its inverse are given as 
 \begin{align} \mathbf U(t)=&\left[\begin{matrix} \omega_1 e^{\lambda_1 t} &0 & \omega_3e^{\lambda_3 t} \\
0 &\Theta(t) &0\\
e^{\lambda_1 t} & 0 &e^{\lambda_3 t}  \end{matrix}\right], \label{FundamentalMatrix}\\ 
{\mathbf U}^{-1}(t)=&\frac{1}{(\omega_1-\omega_3)e^{(\lambda_1+\lambda_3)t}
\Theta(t)}\left[\begin{matrix} \Theta(t) e^{\lambda_3 t} &0 & -\omega_3\Theta(t)e^{\lambda_3 t} \\
0 &e^{(\lambda_1+\lambda_3)t}(\omega_1-\omega_3) &0\\
-\Theta(t) e^{\lambda_1 t} & 0 & \omega_1\Theta(t) e^{\lambda_1 t}  \end{matrix}\right] \label{FundamentalInverse} \end{align}  

\medskip

\noindent{\bf Step 2.\,} Prior to solving the non-homogeneous system \eqref{LS-main} we carry out an additional time-averaging, with the goal of simplifying the problem. Note that the first and second components of the vector field $\mathbf h$, $A_{12}\theta$ and $ A_{21}\phi,$ respectively,  have the structure \, 
$ A_{12}(t)\backsim \frac{L_i}{B}p(t).$
We apply the averaging result that approximates solutions of a system of the form
\be \dot\xx=\epsilon f(t, \xx, \epsilon), \label{averaging-form} \ee 
where $0<\epsilon $ is a small parameter, with those of 
\be \dot\xx=f_{\text{\begin{tiny}{A}\end{tiny}}}(\xx, 0). \ee
Here $ f_{\text{\begin{tiny}{A}\end{tiny}}} $ denotes the function that results from averaging the original $f$ with respect to its explicit $t$-dependence, over an interval that, in our case, corresponds to a full period $[0,1]$. The transformation from our original system \eqref{LS-main} to one of the form \eqref{averaging-form} proceeds by a standard change of variable.   The accuracy of the approximation relies on the smallness of $\epsilon$ \cite{sanders2007averaging}.  

\medskip

\noindent{\bf Step 3.\,} Next, we solve the averaged simplified  non-homogeneous system \eqref{LS-main}.   First, we set up the variation of constant formula that now reads as 
\be \u(t)= \mathbf U(t)\big[{\mathbf U}^{-1}(0)\u(0)+\int_0^t{\mathbf U}^{-1}(s)
\left[\begin{matrix} 0\\0 \\A_{32}(s) q(s)\end{matrix}\right]\,ds\big]. \label{variation-of-parameters-abstract}\ee
Details of the calculation of the terms in the previous equation are shown in the Supplemental Materials section. 
They lead to the solution

\begin{align} 
\phi(t)= &\frac{1}{\omega_1-\omega_3}\big\{ (e^{\lambda_1t}\omega_1 -e^{\lambda_3 t}\omega_3)\phi(0) + \omega_1\omega_3(-e^{\lambda_1t}  +e^{\lambda_3 t})q(0)\nonumber \\+ & \omega_1\omega_3 \int_0^t A_{32}(s)\theta(s)(e^{\lambda_3(t-s)} -e^{-\lambda_1(t-s)})\,ds  \big\}\\
 \theta(t)= & \Theta(t), \\
q(t)= &\frac{1}{\omega_1-\omega_3}\big\{ (e^{\lambda_1 t}-e^{\lambda_3t})\phi(0) +(-\omega_3e^{\lambda_1t} +\omega_1 e^{\lambda_3 t})q(0)\nonumber \\ +& C_2\theta(0)\big(\frac{1}{\lambda_1^2+4\pi^2}[\lambda_1 e^{\lambda_1 t} -\lambda_1\cos{2\pi t} +2\pi\sin{2\pi t}]\nonumber \\ &+\frac{C_1}{2(\lambda_1^2+16\pi^2)} [-\lambda_1\sin{4\pi t} -4\pi\cos{4\pi t} + 4\pi e^{\lambda_1 t}]\big),
\end{align}
with with $\Theta(t)$ as in \eqref{Theta-final}, and  \be C_1= \frac{M\sqrt{c_x}}{4B\pi\Delta_J}, \quad C_2= \frac{2\lambda_\sigma Q_0\Delta_1}{F\Delta_J} \sqrt c_x. \ee
\medskip
\noindent
{\bf Step 4.} We now summarize the neutral stability  conditions that determine the speed and  size range of the soliton-like distortions. These are of two types, involving the selection of  special sets of initial data, and restrictions on the eigenvalues. First of all, from equation \eqref{Theta-final}, we find two cases that lead to related but different instabilities:
\begin{align}
 C_0=&0, \quad    \lambda_3=0\quad  \text {and} \quad \lambda_3-A_{22}^0\neq 0, \quad \text {or}\label{unstability-theta1}\\
 \lambda_3=& A_{22}^0, \quad \text{and}\quad \lambda_3=0. \label{unstability-theta2}
\end{align}
Note that the third condition in \eqref{unstability-theta1} guarantees that $\Theta(0)\neq 0$,  needed for the invertibility of the fundamental matrix solution. Either set of relations also guarantee the preservation of the shape of $\theta(t)$ with time and its phase shift with respect to the applied electric field. Finally, to preserve the shape of $\phi$ and $q$, we additionally require 
\be \lambda_1=0. \label {unstability-q}\ee
The observation in lemma \eqref{lemma3.1} that $A_{13}A_{31}= \frac{4 L_2^2 M Q_0\Delta_\sigma}{BF\Delta_J^2} c_xc_\xi$, with $L_2^2=O(10^{-6})$,  together with 
$\Delta_J\neq 0$, allows us to obtain approximate forms of the vanishing conditions on the eigenvalues $\lambda_1$ and $\lambda_3$ in \eqref{e-values-reduced} as
\be  A_{11}=A_{33} \quad \text{and} \quad A_{11}+ A_{33}=0. \label{threshold-conditions}
\ee
The latter equation provides an expression of the tuxedo speed $V$ as
\be 2V=a_{11}+a_{33}= \frac{1}{\Delta_J}[\frac{4}{B}(C\Delta_1\Delta_J+L_2^2c_xc_\xi)+ \frac{\Delta_\sigma}{F}(4G\Delta_J +Q_0M)]. \label{frequency-and-speed} \ee
The first equation in \eqref{threshold-conditions} evaluated at the parameter values of the problem,  corresponds to the line 
\be c_\xi+0.7878c_x=2.9740.  \label{a11equalsa33-1}\ee 
\begin{remark}
We find that, on the line \eqref{a11equalsa33-1},   $\Delta_\sigma=O(10^{-2}) $  providing the estimate $A_{13}A_{31}=O(L_i^2\times 10^{-2})$ stated after Lemma \eqref{lemma3.1}. Furthermore, $\Delta_1\neq 0$, for $c_x\neq 0$, holds for wave numbers on the line \eqref{a11equalsa33-1}. The later justifies the initial layer argument stated in remark \eqref{initial-layer}.
\end{remark}
Moreover, evaluating $V$ given by \eqref{frequency-and-speed} on the  curve \eqref{a11equalsa33-1}
gives 
\be V=\frac{1}{\Delta_J}[\frac{\Delta_\sigma}{F} (4G\Delta_J+ Q_0 M)]. \label{V-final}\ee

\begin{figure} 
\centering
\includegraphics[height=1.7in]{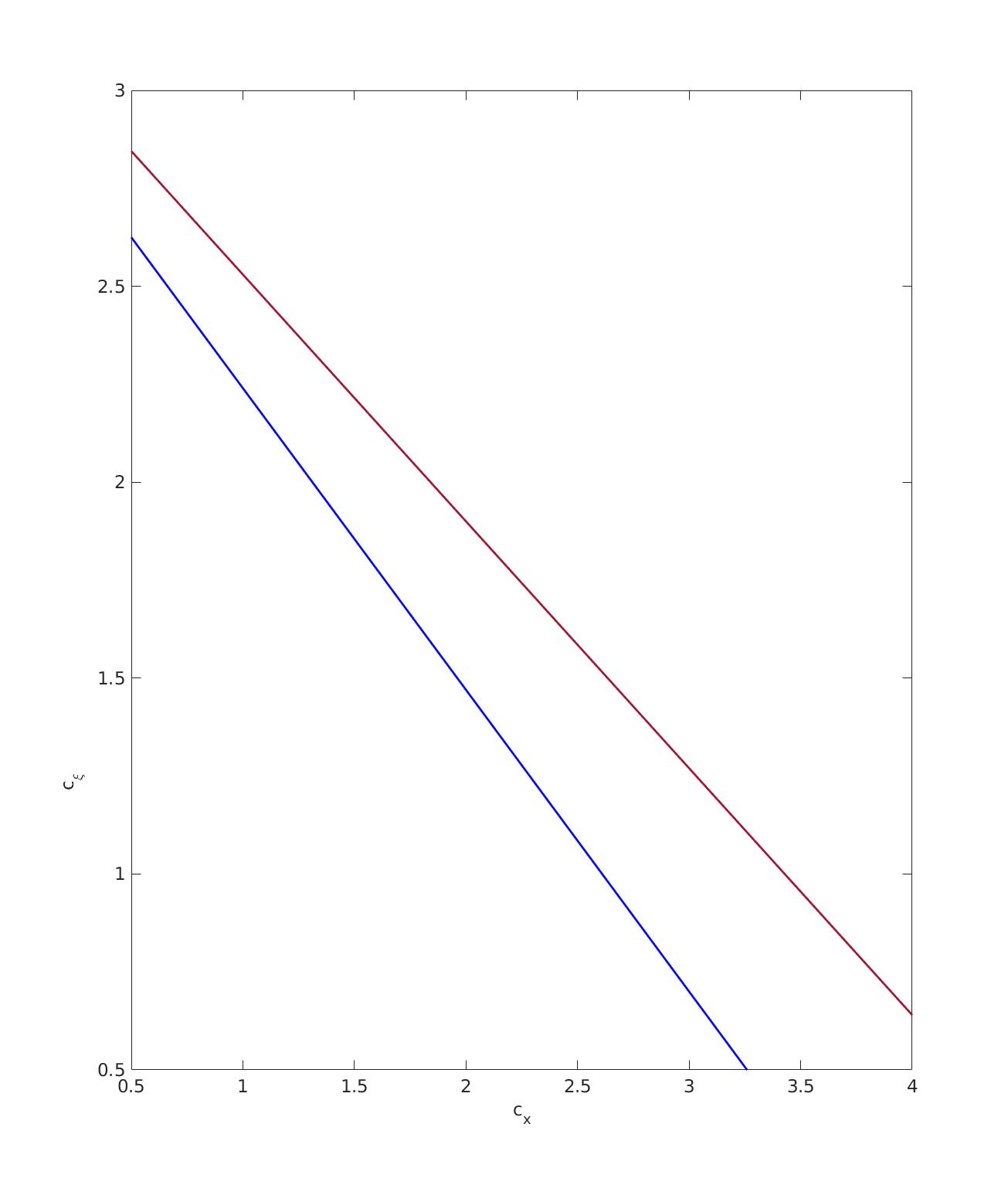}\quad\quad \quad  \includegraphics[height=1.7in]{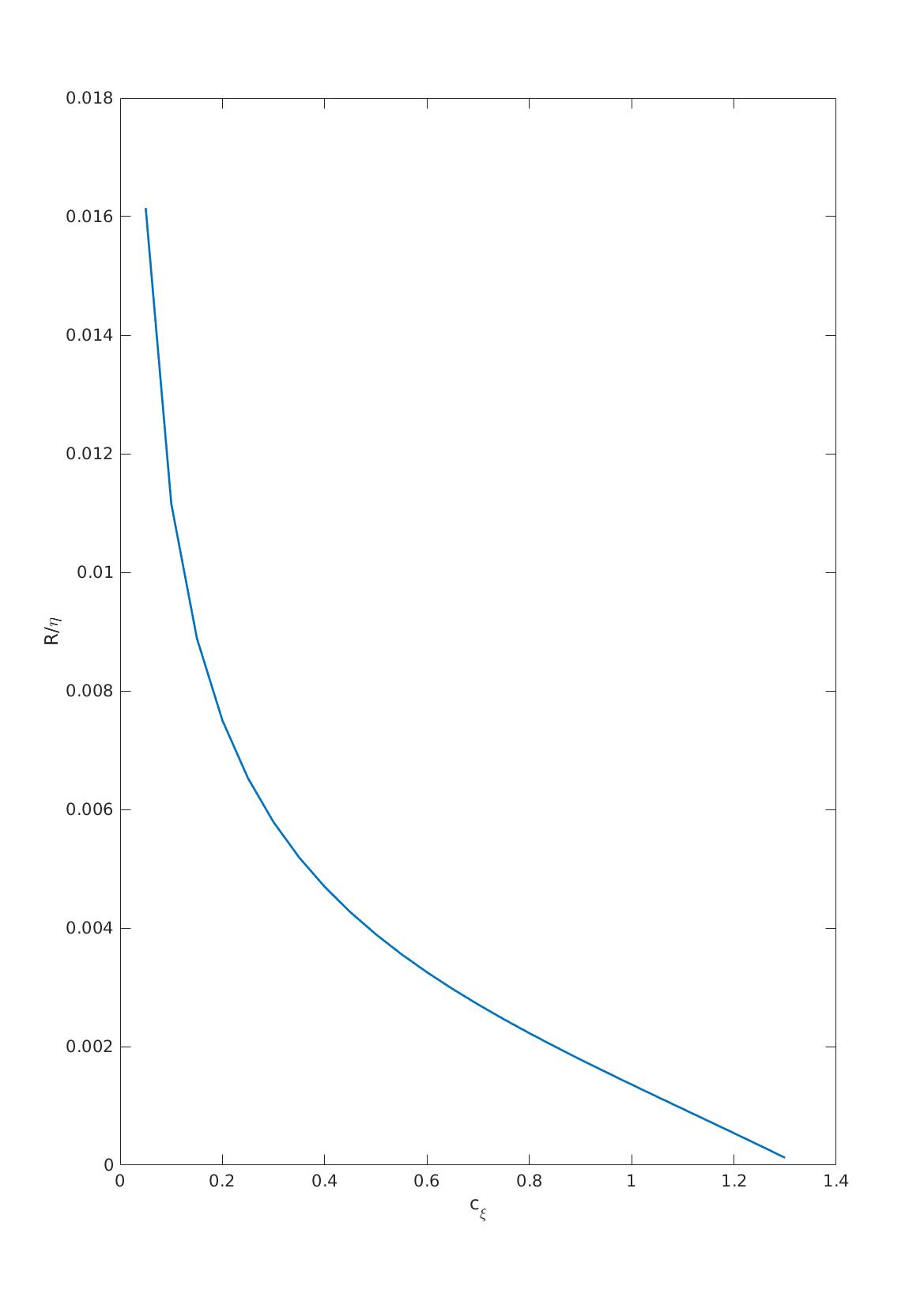}
\caption{The top line on the figure on the left corresponds to the equation $\Delta_J=0$. The bottom line on the same figure is the graph of equation \eqref{a11equalsa33-1}, indicating that $\Delta_J>0$ on the points of the threshold line. The figure on the right represents the  ratio, $\frac{R}{\eta}$, of the dimensionless speed of the soliton-like disturbance over the aspect ratio,  as predicted by the model. The experimental measurements give $\frac{R_{\text{min}}}{\eta}=0.0275 $ and  $\frac{R_{\text{max}}}{\eta}=0.14$ }
\label{final-conditions}
\end{figure}

The expression in \eqref{V-final} indicates that the sign of $V$ is identical to that of $\Delta_\sigma$,  provided $\Delta_J>0.$ We now evaluate $V$ in \eqref{V-final} on the line \eqref{a11equalsa33-1} to get
\be V=\frac{0.0260-0.0127c_x}{F\Delta_J(c_x,c_\xi(c_x))}(4G\Delta_J+ Q_0 M), \label{V-final1}\ee
where the $\Delta_J$ expression in the denominator is evaluated on the line \eqref{a11equalsa33-1}, where it satisfies $\Delta_J>0$ (see also figure \eqref{final-conditions}. 
We observe that 
\be \text{(i)} \, V>0 \,\,  \text {for}\,\,  c_x< 2.0472, \quad V=0 \,\, \text {at}\, \, c_x=2.0472\ \,\, \text {and}\quad  \text{(ii)} \, V<0 \,\, \text {for} \,\, 3.7751\geq c_x>2.0472, \label{V-ranges} \ee where the upper bound on $c_x$ on the region of $V<0$ corresponds to the $c_x$-intercept of the line \eqref{a11equalsa33-1}.
The following lower bound on $c_\xi$ follows from equation \eqref{a11equalsa33-1} and is associated with the $c_x$-interval indicated in \eqref{V-ranges}, for the case $V<0$"
\be 0\leq c_\xi\leq 1.3612:=c_\xi^{\begin{small}\text{max}\end{small}}. \ee
We summarize the latter results on the graphs of Figure \eqref{final-conditions} and obtain estimates for the size of the disturbance.  
First of all, it is immediate to recover the wave numbers $\rho_x$ and $\rho_\xi$ from the quantities $c_x $ and $c_\xi$, and subsequently derive estimates for the tuxedo size, that is, the width $L_x$ and the length $L_\xi$:
\be \begin{aligned}
    |\rho_x|=& \frac{\sqrt{c_x}}{\pi\eta}, \quad  |\rho_\xi|= \frac{\sqrt{c_\xi}}{\pi\eta}, \\
    L_\xi=&|\rho_\xi|^{-1}\geq \frac{\pi\eta}{\sqrt{1.3621}}:= L_\xi^{{\begin{small}\text{min}\end{small}}}=2.6918\eta\\
    \frac{\pi\eta}{\sqrt{2.0472}}=2.1957\eta=&\geq L_x={|\rho_x|}^{-1}\geq \frac{\pi\eta}{\sqrt{3.7751}}=1.6169 \eta.
    \end{aligned} \label{sizes}\ee
We note that the absolute value notation used in the expressions of the wave numbers is consistent with the assumptions \eqref{assume}.
From the estimates in \eqref{sizes}, we see that 
\be   L_{x, {{\begin{small}\text{phys}\end{small}}}}^{{\begin{small}\text{min}\end{small}}} =1.6169 d \quad \text{and}\quad    L_{\xi, {{\begin{small}\text{phys}\end{small}}}}^{{\begin{small}\text{min}\end{small}}} =2.6918 d, \label{sizes-model-prediction-dimensional} \ee 
where the latter quantities denote the dimensional sizes of the tuxedo on the units of the plate gap $d$. 
Comparing the length of the soliton package with that of the electrode plate $L$, we find that 
\be L= 0.33 \times L_{\xi, {{\begin{small}\text{phys}\end{small}}}}^{{\begin{small}\text{min}\end{small}}}\times 10^3. \ee 
From the point of view of the order of magnitude, the previous result agrees with the statement in \cite{Oleg} that the solitons travel on a plate several thousands of their own size. 

We recall that the experimentally measured sizes, the width $L_w$ and length $L_l$, of the soliton-like disturbances are reported as 
\be L_w= 2 d, \quad  20\,\mu\text{m}<L_l< 50\,\mu\text{m}.\ee
Note that, for $d=8\mu\text{m}$, the latter can be written as 
\be 2.5 d\,\mu\text{m}<L_l< 6.25 d \,\mu\text{m}, \ee
showing an excellent agreement with the lower bound \eqref{sizes-model-prediction-dimensional} of the soliton length predicted by our model. 

In particular, the lower bound for the length corresponds to 
\be  c_l= 0.6806=\frac{1}{2}c_\xi^{\begin{small}\text{max}\end{small}}\ee
Next, we evaluate the phase-shift of the out-of-plane angle $\theta$ in \eqref{Theta-final}. For this, 
 we calculate 
\be [2\pi\sin{2\pi t}+ (\lambda_3-A_{22}^0)\cos{2\pi t}]=2\pi [\sin{2\pi t}+ \frac{\lambda_3-A_{22}^0}{2\pi}\cos{2\pi t}]=2\pi \sin({2\pi t-\alpha}), \ee
with \be \sin\alpha\approx \frac{4C}{2\pi B}=O(10^{-2}),  \ee 
where we have applied the second relation in \eqref{unstability-theta1} as well as \eqref{aij}. 

Finally, let us calculate the speed $R$ of the soliton-like packages. First, note that 
\be V=2\pi R i\rho_\xi= -2\pi R \text{Im}\rho_\xi <0 \Longleftrightarrow  R\,\text{Im}\rho_\xi>0.      \ee

Hence, the $c_x$-region such that $V<0$ corresponds to $R>0$, that is, the soliton disturbance traveling along the positive $y$-direction, provided $\text{Im}\rho_x>0$. 
 Furthermore, referring to the formula for the inverse Fourier transform,
we find that $\text{Im }\rho_\xi>0$ corresponds to exponential \emph{decay} along the positive  $y$-direction. 
This, indeed, seems to  correspond to the realistic physical setting, indicating that  the perturbation has not yet reached, at time t>0, locations such that $y>> Rt$. 

Let us now compare the experimentally measured speed range $R$ in \eqref{dimensionless-soliton-speed-range} with the values displayed in the graph in figure \eqref{final-conditions}. We first note that, for the aspect ratio  $\eta=1.6\times 10^{-3}$,
the  experimental range  previously referred to yields
\be 0.0275\leq \frac{R}{\eta}\leq 0.1400. \label{experimental-range-scaled} \ee 
The discussion on the predicted soliton-like sizes in \eqref{sizes} suggests that the corresponding  speeds, as shown in the graph of $R-c_\xi$ in figure \eqref{final-conditions}, fit towards the lower bound indicated in \eqref{experimental-range-scaled}. If smaller solitons than the latter ones are to be taken into account, they would travel at speeds below than the lower bound in \eqref{experimental-range-scaled}, up to an order of magnitude smaller. In conclusion, a decrease of the inter-plate gap $d$ would bring a tighter agreement with the experimental findings. 

We discard  the solutions corresponding to $V>0$ in \eqref{a11equalsa33-1}  that would yield disturbances with an exceedingly large  horizontal size,  vertically too short, and also with an incompatible growth rate. 

\section{Acknowledgment} 
The authors wish to gratefully knowledge the support of the National Science Foundation, through the grant DMS-DMREF 1729589. They also want to express their gratitude to Professor Oleg Lavrentovich for the many discussions and the sharing of experimental results, and to Professor Dmitry Golovaty for his helpful comments.
\section{Conclusion}
In this article, we have develop a time dependent model of a flexoelectric nematic liquid crystal that couples elastic, viscous, conducting, dielectric and flexoelectric effects. We use linear analysis to investigate the three-dimensional solitons observed when such a material is subject to an alternating electric field, within the appropriate range of intensity and frequency.  The work focuses on finding the instability threshold of the uniformly aligned nematic, and yields estimates on the size, phase-shift and speed of the soliton-like package. The length and speed of the soliton predicted by the model fall towards the lower range of the experimentally measured ones. The work presented here is the first part of the three article set devoted to the study of the physical solitons reported in \cite{Oleg}.
A forthcoming nonlinear analysis aims at correcting the lower predictions of the present linear model, that will also address finer aspects of the soliton shape such as the size of the head versus the tail.  Finally, the third article will focus on the well-posedness of flexoelectric nematic models, that present additional nontrivial challenges due to the higher order gradient of the theory, compared with standard nematic.

\bibliographystyle{unsrt}
\bibliography{thesis_bib}


\end{document}


\maketitle
\section{Supplemental material} 
\subsection{Details of the solution of the non-homogeneous system by the variation of constants formula}
Next, we solve the averaged simplified  non-homogeneous system, equation (3-25) of the article.   First, we set up the variation of constant formula that now reads as 
\be \u(t)= \mathbf U(t)\big[{\mathbf U}^{-1}(0)\u(0)+\int_0^t{\mathbf U}^{-1}(s)
\left[\begin{matrix} 0\\0 \\A_{32}(s) q(s)\end{matrix}\right]\,ds\big].\ee
Let us calculate the terms involved in the previous vector equation:

\be {\mathbf U}^{-1}(0)= \frac{1}{(\omega_1-\omega_3)
\Theta(0)}\left[\begin{matrix} \Theta(0)  &0 & -\omega_3\Theta(0) \\
0 &(\omega_1-\omega_3) &0\\
-\Theta(0) & 0 & \omega_1\Theta(0) \end{matrix}\right]  \ee

\be  {\mathbf U}(t)  {\mathbf U}^{-1}(0) = \frac{1}{(\omega_1-\omega_3)\Theta(0)}\left[\begin{matrix} \Theta(0)(e^{\lambda_1 t}\omega_1-e^{\lambda_3t}\omega_3) & 0 &\Theta(0)\omega_1\omega_3(-e^{\lambda_1t}+ e^{\lambda_3 t})\\ 0 &\Theta(t)(\omega_1-\omega_3) & 0\\\Theta(0)(e^{\lambda_1 t}-e^{\lambda_3t}) & 0&\Theta(0)(e^{\lambda_3 t}\omega_1-e^{\lambda_1t}\omega_3) \end{matrix}\right] \nonumber
\ee

\be 
 {\mathbf U}(t)  {\mathbf U}^{-1}(0)\u_0(0)= \frac{1}{\omega_1-\omega_3} \left[\begin{matrix} (e^{\lambda_1t}\omega_1 -e^{\lambda_3 t}\omega_3)\phi(0) + \omega_1\omega_3(-e^{\lambda_1t}  +e^{\lambda_3 t})q(0)\\
 \theta(0)\Theta(t)\Theta(0)^{-1}(\omega_1-\omega_3)\\
 (e^{\lambda_1 t}-e^{\lambda_3t})\phi(0) +(-\omega_3e^{\lambda_1t} +\omega_1 e^{\lambda_3 t})q(0)\end{matrix}\right]
\ee
\be  {\mathbf U}^{-1}(s)\left[\begin{matrix} 0\\0\\A_{32}\theta(s)\end{matrix}\right]= \frac{1}{\omega_1-\omega_3}
\left[\begin{matrix} -\omega_3 e^{-\lambda_1 s}A_{32}(s)\theta(s) \\ 0\\ \omega_1e^{-\lambda_3 s}A_{32}(s)\theta(s)\end{matrix}\right] \ee

\be \mathbf U(t) \int_0^t {\mathbf U}^{-1}(s)\left[\begin{matrix} 0\\0\\A_{32}\theta(s)\end{matrix}\right]\,ds =
\frac{1}{\omega_1-\omega_3}\left[\begin{matrix} \omega_1\omega_3\int_0^t A_{32}(s)\theta(s)(e^{\lambda_3(t-s)} -e^{-\lambda_1(t-s)})\,ds \\0 \\ \int_0^t A_{32}(s)\theta(s)(\omega_1e^{\lambda_3(t-s)}-\omega_3 e^{\lambda_1(t-s)})\,ds \end{matrix}\right]\ee
Finally, we have
\be\begin{aligned}
\u(t)=   \frac{1}{\omega_1-\omega_3}&\{ \left[\begin{matrix} (e^{\lambda_1t}\omega_1 -e^{\lambda_3 t}\omega_3)\phi(0) + \omega_1\omega_3(-e^{\lambda_1t}  +e^{\lambda_3 t})q(0)\\
 \theta(0)\Theta(t)\Theta(0)^{-1}(\omega_1-\omega_3)\\
 (e^{\lambda_1 t}-e^{\lambda_3t})\phi(0) +(-\omega_3e^{\lambda_1t} +\omega_1 e^{\lambda_3 t})q(0)\end{matrix}\right] \\
 &+ \left[\begin{matrix} \omega_1\omega_3\int_0^t A_{32}(s)\theta(s)(e^{\lambda_3(t-s)} -e^{-\lambda_1(t-s)})\,ds \\0 \\ \int_0^t A_{32}(s)\theta(s)(\omega_1e^{\lambda_3(t-s)}-\omega_3 e^{\lambda_1(t-s)})\,ds \end{matrix}\right]\} \label{u-soln}
\end{aligned}\ee 
In particular, we see that 
\be \theta(t)= {\theta(0)} \Theta(t), \ee
with $\Theta(t)$ as in equation (3.41) of the article. Next, we calculate
\be \int_0^t A_{32}(s)\theta(s) e^{-\lambda_i s}\,ds =\theta(0)   \int_0^t A_{32}(s)\Theta(s) e^{-\lambda_i s}\,ds, \quad i=1, 3  \ee
Let us calculate
\begin{align*}
   \int_0^t& A_{32}(s)\Theta(s) e^{\lambda_1(t- s)}\,ds  = C_2  \int_0^t \cos{2\pi s}(1+ C_1\sin{2\pi s}) e^{\lambda_1(t-s)}\,ds\nonumber\\ &= C_2\big(\frac{1}{\lambda_1^2+4\pi^2}[\lambda_1 e^{\lambda_1 t} -\lambda_1\cos{2\pi t} +2\pi\sin{2\pi t}] +\frac{C_1}{2(\lambda_1^2+16\pi^2)} [-\lambda_1\sin{4\pi t} -4\pi\cos{4\pi t} + 4\pi e^{\lambda_1 t}]\big),
\end{align*}
where \be C_1= \frac{M\sqrt{c_x}}{4B\pi\Delta_J}, \quad C_2= \frac{2\lambda_\sigma Q_0\Delta_1}{F\Delta_J} \sqrt c_x. \ee
The first component of \eqref{u-soln} is
\begin{align} 
\phi(t)= &\frac{1}{\omega_1-\omega_3}\big\{ (e^{\lambda_1t}\omega_1 -e^{\lambda_3 t}\omega_3)\phi(0) + \omega_1\omega_3(-e^{\lambda_1t}  +e^{\lambda_3 t})q(0)\nonumber \\+ & \omega_1\omega_3 \int_0^t A_{32}(s)\theta(s)(e^{\lambda_3(t-s)} -e^{-\lambda_1(t-s)})\,ds  \big\}
\end{align}

The third component of \eqref{u-soln} is 
\begin{align}
q(t)= &\frac{1}{\omega_1-\omega_3}\big\{ (e^{\lambda_1 t}-e^{\lambda_3t})\phi(0) +(-\omega_3e^{\lambda_1t} +\omega_1 e^{\lambda_3 t})q(0)\nonumber \\ +& C_2\theta(0)\big(\frac{1}{\lambda_1^2+4\pi^2}[\lambda_1 e^{\lambda_1 t} -\lambda_1\cos{2\pi t} +2\pi\sin{2\pi t}]\nonumber \\ &+\frac{C_1}{2(\lambda_1^2+16\pi^2)} [-\lambda_1\sin{4\pi t} -4\pi\cos{4\pi t} + 4\pi e^{\lambda_1 t}]\big)
\end{align}